\documentclass[11pt,a4paper]{article}

\usepackage{subfigure,jcappub,bm}

\title{Generation of the magnetic helicity in a neutron star driven by the electroweak electron-nucleon interaction}

\author[a,b,c]{Maxim Dvornikov}
\author[b]{Victor B. Semikoz}
\affiliation[a]{Institute of Physics, University of S\~{a}o Paulo, CP 66318, CEP 05314-970 S\~{a}o Paulo, SP, Brazil}
\affiliation[b]{Pushkov Institute of Terrestrial Magnetism, Ionosphere
and Radiowave Propagation (IZMIRAN),
142190 Troitsk, Moscow, Russia}
\affiliation[c]{Physics Faculty, National Research Tomsk State University,
36 Lenin Ave., 634050 Tomsk, Russia}
\emailAdd{maxdvo@izmiran.ru}
\emailAdd{semikoz@yandex.ru}

\abstract{
We study the instability of magnetic fields in a neutron star core driven by the parity violating part of the electron-nucleon interaction in the Standard Model. Assuming a seed field of the order $10^{12}\thinspace\text{G}$, that is a common value for pulsars, one obtains its amplification due to such a novel mechanism by about five orders of magnitude, up to $10^{17}\thinspace\text{G}$, at time scales $\sim (10^3 - 10^5)\thinspace\text{yr}$. This effect is suggested to be a possible explanation of the origin of the strongest magnetic fields observed in magnetars. The growth of a seed magnetic field energy density
is stipulated by the corresponding growth of the magnetic helicity density 
due to the presence of the anomalous electric current in the Maxwell equation.  Such an anomaly is the sum of the two competitive effects: (i) the chiral magnetic effect driven by the difference of chemical potentials for the right and left handed massless electrons
and (ii) constant chiral electroweak electron-nucleon interaction term, 
which has the polarization origin and depends on the constant neutron density in a neutron star core. The remarkable issue for the decisive role of the magnetic helicity evolution in the suggested mechanism is the arbitrariness of an initial magnetic helicity including the case of non-helical fields from the beginning.
The tendency of the magnetic helicity density to the maximal helicity case at large evolution times provides the growth of a seed magnetic field	to the strongest magnetic fields in astrophysics.
}


\begin{document}

\maketitle

\section{Introduction}

The study of strong magnetic fields inherent in some compact astrophysical objects, like neutron stars, is one of the hottest topics of the modern astrophysics~\cite{LatPra04}. This research received a significant impact in the wake of the observations of anomalous X-ray pulsars (AXP)~\cite{Maz79} and soft gamma-ray repeaters (SGR)~\cite{FahGre81}. Some observational characteristics of SGRs and AXPs distinguish them from the more common accretion-powered pulsars in massive X-ray binaries and place them into a separate class of astrophysical objects.

From the point of view of modern astrophysics~\cite{Mer08}, SGRs and AXPs are supposed to be highly magnetized, $B \gtrsim 10^{15}\thinspace\text{G}$, neutron stars, or magnetars. It should be noted that a typical pulsar can have a magnetic field up to $B_0 \sim 10^{12}\thinspace\text{G}$. The magnetic field  $\sim B_0$ can be generated in the core of a supernova (SN) progenitor at some earlier stage of its evolution, and subsequently amplified during core collapse, for example, simply by the flux conservation \cite{Spruit}. The source of the strong magnetic field in a magnetar, in its turn, is still unclear.

It was shown in refs.~\cite{AllHor04,VinKui06} that the most popular model of the magnetic field in a magnetar, based on the turbulent dynamo~\cite{Duncan}, confronts with some of the observational data, and thus should be corrected. Some other models based, e.g., on the idea of the fossil field, are also discussed in the literature~\cite{VinKui06}.

Anyway, if one adopts a dynamo scenario for the creation of the magnetic field in a magnetar, one has to propose a mechanism to amplify the seed field $B_0 = 10^{12}\thinspace\text{G}$, which is typical for a pulsar, by at least three orders of magnitude. Recently, in ref.~\cite{Dvornikov:2014uza}, we put forward a new mechanism which can provide this kind of amplification. That scenario is based on the magnetic field instability because of the parity violation in the Standard Model (SM) for the electroweak electron-nucleon ($eN$) interaction of ultrarelativistic degenerate electrons with non-relativistic degenerate nucleons in a neutron star (NS) as the SN remnant and a future magnetar.

On the first glance the mechanism proposed in ref.~\cite{Dvornikov:2014uza} resembles the results of ref.~\cite{Vilenkin}, where the chiral magnetic effect (CME) in an external magnetic field, which is proportional to the chemical potential of {\it massless} charged fermions, was established. The contribution of the electroweak interaction into the averaged electron current was discarded in the case of {\it massive} fermions. That result of ref.~\cite{Vilenkin} is based on the ambiguous renormalization procedure for the photon polarization operator calculated in the first order of the perturbation theory in $G_\mathrm{F}$, where  $G_\mathrm{F} \approx 1.17 \times 10^{-5}\thinspace\text{GeV}^{-2}$ is the Fermi constant. Nevertheless, the possibility that a non-zero weak interaction term can appear in other models has not been ruled out in ref.~\cite{Vilenkin}.

To avoid nonsequential method in ref.~\cite{Vilenkin} during calculations of the electric current driven by electroweak interactions, in ref.~\cite{Dvornikov:2014uza} we considered the additive $eN$ interaction in the same Dirac equation for {\it chiral states} of  \emph{massless} electrons in an external uniform magnetic field ${\bf B}$. We obtained that the total induced electric current ${\bf J}$, which is additive to the standard ohmic current ${\bf J}_\mathrm{Ohm}$ in plasma, entering the Maxwell equation,
\begin{equation}\label{current}
  {\bf J} = \frac{2\alpha_\mathrm{em}}{\pi}(\mu_5 + V_5){\bf B},
\end{equation}
is given by the difference of the right and left electrons chemical potentials, $\mu_5=(\mu_\mathrm{R} - \mu_\mathrm{L})/2$, and includes the non-zero weak interaction coefficient
\begin{equation}\label{weak}
  V_5 = \frac{(V_\mathrm{L} - V_\mathrm{R})}{2} =
  \frac{G_\mathrm{F}}{2\sqrt{2}}
  \left[
    n_n - (1 - 4\xi)n_p
  \right],
\end{equation}
where $n_{n,p}$ are the densities of neutrons and protons inside NS, $\xi=\sin^2\theta_\mathrm{W}=0.23$ is the Weinberg parameter and $\alpha_\mathrm{em} \approx \mathrm (137)^{-1}$ is the fine structure constant in eq.~(\ref{current}).

We would like to stress the important role of the magnetic helicity density $h(t)=H/V=V^{-1}\int (\mathbf{A}\cdot\mathbf{B}){\rm d}^3x$ in the generation of magnetic fields ${\bf B}=\nabla\times {\bf A}$ in magnetars\footnote{It is well-known that, in a perfectly conducting fluid with fixed boundary conditions, the magnetic helicity $H$ is a conserved quantity. Even for a finite electric conductivity the magnetic helicity changes slowly. It leads to the existence of {\it stable equilibrium configurations} for magnetic fields. To the extent that the helicity is conserved, a non-equilibrium or unstable magnetic field with a finite conductivity cannot decay completely, since the helicity of a vanishing field is zero.}.

Despite the problem of the magnetic field evolution in a SN remnant was treated in a simplified way in ref.~\cite{Dvornikov:2014uza}, we could derive some of the characteristics of the magnetic field in a magnetar, which are close to the observed ones. Nevertheless we should mention some of the shortcomings of our model in ref.~\cite{Dvornikov:2014uza}. Firstly, a stationary electric conductivity $\sigma_\mathrm{cond}=\text{const} \sim (T_8)^{-2}$ at the typical temperature $T\simeq 10^8\thinspace\text{K}$ for a cooling neutron star instead of its decreasing temperature $T_8(t)=T(t)/10^8\thinspace\text{K}$ was used in the analysis.
Secondly, we assumed the maximal and {\it monochromatic} helicity density spectrum: $h(k,t)=2\rho_\mathrm{B}(k,t)/k_0$ and $h(k,t)=h(t)\delta (k - k_0)$, where $\rho_\mathrm{B}(k,t)$ is the spectrum of the magnetic energy density. Thirdly, we considered the magnetic field with the largest scale $\Lambda_\mathrm{B}=k_0^{-1}=R_\mathrm{NS}=10\thinspace\text{km}$ only.

Of course, the canonical inequality $h(k,t)\leq 2\rho_\mathrm{B}(k,t)/k$ (see, e.g., ref.~\cite{Biscamp}) with a running wave number $k$, $k_\mathrm{min}\leq k\leq k_\mathrm{max}$, for the {\it continuous} helicity density spectrum $h(k,t)$ and the magnetic energy density spectrum $\rho_\mathrm{B}(k,t)$, as well as the diminishing temperature for a cooling NS core, $\mathrm{d}T/\mathrm{d}t<0$, are more realistic conditions for our MagnetoHydroDynamical (MHD) problem, and we can improve our model in ref.~\cite{Dvornikov:2014uza} using all these ingredients.

Our work is organized as follows. In the main section~\ref{MAGNHELEVOL} we consider the magnetic helicity evolution in a cooling NS core starting from the Faraday equation generalized in SM due to the anomalous electric current in eq.~(\ref{current}). Basing on the Faraday equation and using the Adler anomaly, we derive in section~\ref{MAGNHELEVOL} the full set of self-consistent kinetic equations for the three functions: the magnetic helicity density spectrum $h(k,t)$, the magnetic energy density spectrum $\rho_\mathrm{B}(k,t)$ and the  CME parameter $\mu_5(t)$. After formulation of the initial conditions in section~\ref{init-condition}, we explain in section~\ref{t0} our choice of the initial time for which the law of the temperature cooling within a NS core should be valid. Then we present our results with the numerical solutions of the three kinetic equations for the magnetic helicity density $h(t)$ in section~\ref{helicity-growth}, for the CME parameter $\mu_5(t)$ in section~\ref{SINGCME}, and, finally, the magnetic field $B(t)$ in the separate section~\ref{BGEN}. In section~\ref{CONCL} we discuss our results and comment on some remaining problems. The detailed derivation of the induced electric current in eq.~\eqref{current} is given in appendix~\ref{CURRDER}.

\section{Magnetic helicity evolution in a cooling NS core\label{MAGNHELEVOL}}

In this section we shall derive the system of kinetic equations for the continuous spectra of the helicity density and the magnetic energy density as well as for the chiral imbalance in the presence of the electroweak term $V_5$ in the current in eq.~(\ref{current}). We rewrite this system using the dimensionless variables in the form convenient for the numerical simulation, and obtain its solutions which obey certain initial conditions.

We shall start with the Faraday equation for the magnetic field $\mathbf{B}$ evolution,
\begin{equation}\label{Faraday}
  \frac{\partial {\bf B}}{\partial t} =
  \alpha (t)\nabla\times {\bf B} +  \eta (t)\nabla^2{\bf B},
\end{equation}
which can be obtained in a standard way from the Maxwell equations assuming the MHD approximation; cf. in refs.~\cite{Moffat,ZelRuzSok90}. Here $\alpha (t)=\Pi (t)/\sigma_\mathrm{cond}$ is the magnetic helicity parameter given by the anomalous current in eq.~(\ref{current}) with the coefficient
\begin{equation}\label{helicity}
  \Pi (t)=\frac{2\alpha_\mathrm{em}}{\pi}(\mu_5(t) + V_5),
\end{equation}
and $\eta (t)=(\sigma_\mathrm{cond})^{-1}$ is the magnetic diffusion coefficient given by the electric conductivity $\sigma_\mathrm{cond}$. Note that one has to take into account the time dependence of $\sigma_\mathrm{cond} = \sigma_\mathrm{cond}(t)$ in the cooling NS core; cf. ref.~\cite{Kelly}. In deriving eq.~\eqref{Faraday}, we suppose that the anomalous current in eq.~\eqref{current} is added to the standard ohmic current ${\bf J}_\mathrm{Ohm}=\sigma_\mathrm{cond}({\bf E} + {\bf v}\times {\bf B})$ \footnote{Throughout the text we have neglected the bulk velocity evolution described by the Navier-Stokes equation since the length scale of the velocity variation $\lambda_v$ is much shorter than the correlation distance of the magnetic field, $\lambda_v\ll k^{-1}$. In other words, infrared modes of the magnetic field  are practically unaffected by the velocity of plasma. In addition, the bulk velocity ${\bf v}$ does not contribute to the helicity evolution $\mathrm{d}h/\mathrm{d}t\sim ({\bf E}\cdot{\bf B})$ when the generalized Ohm law is substituted, ${\bf E}= - {\bf v}\times {\bf B} + \eta\nabla\times {\bf B} - \alpha{\bf B}$. The small scale $\lambda_v$ is also a reason why we omitted the dynamo term $\nabla \times ({\bf v}\times {\bf B})$ in the Faraday equation. Such a term could be important for a turbulent (early) stage of the NS evolution accounting for its differential rotation.}.

Basing on eq.~\eqref{Faraday}, one can write the evolution equations for the binary combinations, $h(t)\sim AB$ and $\rho_\mathrm{B}(t)\sim B^2$. Here $h(t)$ is the magnetic helicity density,
\begin{equation}\label{hdef}
  h(t) = \frac{1}{V} \int \mathrm{d}^3x ({\bf A}\cdot{\bf B}) =
  \frac{1}{V} \int \frac{\mathrm{d}^3k}{(2\pi)^3} ({\bf A}_k\cdot{\bf B}^*_k) =
  \int h(k,t) \mathrm{d}k,
\end{equation}
where $V$ is the normalization volume, $h(k,t)$ is the spectrum of the helicity density, ${\bf A}_k$ and ${\bf B}_k$ are the Fourier components of the vector potential ${\bf A}$ and the magnetic field ${\bf B}$.  The equation for $h(k,t)$ should be complemented by the equation for the magnetic energy density $\rho_\mathrm{B}(t)$,
\begin{equation}\label{rhodef}
  \rho_\mathrm{B}(t) = \frac{1}{2V} \int \frac{\mathrm{d}^3k}{(2\pi)^3} |{\bf B}_k|^2 =
  \int \mathrm{d}k \rho_\mathrm{B}(k,t)=\frac{1}{2}B^2(t).
\end{equation}
Note that, in eqs.~\eqref{hdef} and~\eqref{rhodef}, while defining $h(k,t)$ and $\rho_\mathrm{B}(k,t)$,
\begin{equation}\label{spectra}
  h(k,t) = \frac{k^2}{2\pi^2V}{\bf A}(k,t)\cdot{\bf B}^*(k,t),
  \quad
  \rho_\mathrm{B}(k,t) = \frac{k^2}{4\pi^2V}{\bf B}(k,t)\cdot{\bf B}^*(k,t),
\end{equation}
we perform the integration over the angles in the Fourier space meaning isotropic spectra.

Accounting for the definition of $h(k,t)$ and $\rho_\mathrm{B}(k,t)$ in eq.~\eqref{spectra}, their evolution reads $\partial_t h(k,t)\sim (\dot{{\bf A}}_k\cdot{\bf B}_k^* +{\bf A}_k\cdot\dot{{\bf B}}^*_k)$ and $\partial_t\rho_\mathrm{B}(k,t)\sim (\dot{{\bf B}}_k\cdot{\bf B}_k^* + \dot{{\bf B}}_k^*\cdot{{\bf B}}_k)$. Finally one gets that (see also Appendix~D in ref.~\cite{Dvornikov:2013bca}):
\begin{eqnarray}\label{general}
  &&\frac{\partial h(k,t)}{\partial t} =
  -\frac{2k^2}{\sigma_\mathrm{cond}}h(k,t) +
  \left(
    \frac{4\Pi}{\sigma_\mathrm{cond}}
  \right)
  \rho_\mathrm{B}(k, t),
  \nonumber\\&&
  \frac{\partial \rho_\mathrm{B}(k,t)}{\partial t}=
  -\frac{2k^2}{\sigma_\mathrm{cond}}\rho_\mathrm{B}(k,t) +
  \left(
    \frac{\Pi}{\sigma_\mathrm{cond}}
  \right)
  k^2 h(k, t).
\end{eqnarray}
where $\Pi = \Pi(t)$ is the parameter in eq.~(\ref{helicity}) responsible for the magnetic field instability. Note that analogous evolution equations were obtained in ref.~\cite{Cam07}.

Then we should derive the kinetic equation which governs the chiral imbalance $\mu_5(t)$ being complementary to the system in eq.~(\ref{general}). For that purpose we use the Adler anomaly~\cite{Adler}, $\partial_{\mu}(j^{\mu}_\mathrm{R} -j^{\mu}_\mathrm{L})=\partial_{\mu}(\bar{\psi}\gamma^{\mu}\gamma^5\psi)= (2\alpha_\mathrm{em}/\pi)({\bf E}\cdot{\bf B})$, and the magnetic helicity density evolution
\begin{equation}\label{change}
  \frac{{\rm d}h(t)}{{\rm d}t} =
  -\frac{2}{V}\int \mathrm{d}^3x({\bf E}\cdot{\bf B}),
\end{equation}
which results from the Maxwell equations and eq.~\eqref{hdef}. Integrating the Adler anomaly, $\tfrac{1}{V}\int \mathrm{d}^3x(...)$, combined with eq~(\ref{change}), one gets the conservation law
\begin{equation}\label{conservation}
  \frac{{\rm d}}{{\rm d}t}
  \left[
    n_\mathrm{R} - n_\mathrm{L} +
    \frac{\alpha_\mathrm{em}}{\pi}h(t)
  \right]=0,
\end{equation}
where $n_\mathrm{R,L}$ are the electron densities for right and left handed electrons given by their chemical potentials $\mu_\mathrm{R,L}$.

Note that at the beginning of a chiral imbalance, $\mu_\mathrm{R}\sim \mu_\mathrm{L}\sim \mu$. Here $\mu$ is the chemical potential of the ultrarelativistic degenerate electron gas which is fixed for the conventional abundance of electrons $Y_e\approx 0.05$ in a cooling NS. For the nucleon density close to the nuclear one, $n_\mathrm{B}\approx n_n=0.18\thinspace\mathrm{fm}^{-3}$, the abundance $Y_e=n_e/n_\mathrm{B}=0.05$ corresponds to the electron density $n_e=\mu^3/3\pi^2=9\times 10^{36}\thinspace\text{cm}^{-3}$, or $\mu=125\thinspace\text{MeV}=\text{const}$.

Using the expression for the derivative of the density difference $\mathrm{d} (n_\mathrm{R} - n_\mathrm{L}) /\mathrm{d} t \approx 2\dot{\mu}_5(t)\mu^2/\pi^2$ and the first line in eq.~(\ref{general}) for ${\rm d}h(t)/{\rm d}t= \int \mathrm{d}k [\partial h(k,t)/\partial t]$, the conservation law in eq.~(\ref{conservation}) can be rewritten as the kinetic equation for the imbalance $\mu_5(t)$,
\begin{equation}\label{kinetics2}
  \frac{\mathrm{d}\mu_5(t)}{\mathrm{d}t}=
  \frac{\pi\alpha_\mathrm{em}}{\mu^2 \sigma_\mathrm{cond}}
  \int \mathrm{d} k \thinspace k^2h(k,t)
  -
  \left[
    \frac{4\alpha_\mathrm{em}^2\rho_\mathrm{B}(t)}{\mu^2\sigma_\mathrm{cond}}
  \right]
  (\mu_5(t) + V_5) - \Gamma_f\mu_5,
\end{equation}
where we added the rate of chirality-flipping processes, $\Gamma_f\approx (m_e/\mu)^2\nu_\mathrm{coll}$, given by the Rutherford electron-proton ($ep$) collision frequency $\nu_\mathrm{coll}=\omega_p^2/\sigma_\mathrm{cond}$ without flip. Here $\omega_p = \mu \sqrt{4\alpha_\mathrm{em}/3\pi}$  is the plasma frequency in a degenerate ultrarelativistic electron gas. Note that the value of $\sigma_\mathrm{cond}$ should correspond to a degenerate electron-proton plasma consisting of ultrarelativistic degenerate electrons and nonrelativistic degenerate protons.

Let us introduce the notations for the dimensionless functions:
\begin{equation}\label{notations}
  \mathcal{H}(\kappa,\tau)=\frac{\alpha_\mathrm{em}^2}{2\mu^2}h(k,t),
  \quad
  \mathcal{R}(\kappa,\tau)=
  \frac{\alpha_\mathrm{em}^2}{k_\mathrm{min}\mu^2}\rho_\mathrm{B}(k,t),
  \quad
  \mathcal{M}(\tau)=\frac{\alpha_\mathrm{em}}{\pi k_\mathrm{min}}\mu_5(t),
\end{equation}
where $k_\mathrm{min}$ is the lower bound for the wave number $k$ range (see below). Then, using eqs.~\eqref{general} and~\eqref{kinetics2}, we get
\begin{eqnarray}\label{system}
  && \frac{\partial \mathcal{H}(\kappa,\tau)}{\partial \tau} =
  F
  \left[
    -\kappa^2\mathcal{H}(\kappa,\tau) +
    2(\mathcal{M}(\tau) + \mathcal{V})
    \mathcal{R}(\kappa,\tau)
  \right],
  \nonumber
  \\
  && \frac{\partial \mathcal{R}(\kappa,\tau)}{\partial \tau} =
  F
  \left[
    -\kappa^2\mathcal{R}(\kappa,\tau) +
    2(\mathcal{M}(\tau) + \mathcal{V})
    \kappa^2\mathcal{H}(\kappa,\tau)
  \right],
  \nonumber
  \\
  && \frac{{\rm d}\mathcal{M}(\tau)}{{\rm d}\tau} =
  F
  \left[
    \int^{\kappa_\mathrm{max}}_1
    \kappa^2\mathcal{H}(\kappa,\tau)\mathrm{d}\kappa -
    2(\mathcal{M}(\tau) + \mathcal{V})
    \int^{\kappa_\mathrm{max}}_1\mathcal{R}(\kappa,\tau)\mathrm{d}\kappa -
    \mathcal{G}\mathcal{M}(\tau)
  \right].
\end{eqnarray}
Here the argument $\kappa=k/k_\mathrm{min}$ runs in the wave number region $k_\mathrm{min}=R_\mathrm{NS}^{-1}\leq k\leq k_\mathrm{max}$, $\tau=2k_\mathrm{min}^2t/\sigma_0$ is the dimensionless diffusion time, $ \mathcal{G}=(\sigma_0\Gamma_f/2k_\mathrm{min}^2)/F=(2\alpha_\mathrm{em}/3\pi)(m_e/k_\mathrm{min})^2$ is the dimensionless rate of the chirality flip,
and $\sigma_0$ is the electric conductivity at the initial time $t_0$ when the temperature of a NS core was $T_0=10^8\thinspace\text{K}$.
The factor $F(\tau)=\sigma_0/\sigma_\mathrm{cond}(t)=[T(\tau)/T_0]^2$ characterizes an increase of the electric conductivity $\sigma_\mathrm{cond}(t)\sim T^{-2}$ during the cooling of a NS core due to the neutrino (antineutrino) emission (see below). Finally, $\mathcal{V}=\alpha_\mathrm{em}V_5/\pi k_\mathrm{min}=7\times 10^8$ is the dimensionless $eN$ weak interaction potential for the fixed $V_5=6\thinspace\text{eV}$ in eq.~(\ref{weak}) given by the constant neutron density, $n_n\gg n_p$, where $n_n=0.18\thinspace\text{fm}^{-3}$.

Note that the system in eq.~\eqref{system} is a generalization of the kinetic equations derived in ref.~\cite{Dvornikov:2014uza} for the case of an {\it arbitrary helicity density}, based on the system of the two kinetic eqs.~(\ref{general}) completed by the evolution of the chiral imbalance $\mu_5(t)$ in eq.~(\ref{kinetics2}). Indeed, considering the particular case of the monocromatic helicity density spectrum, $\mathcal{H}(\kappa,\tau)=\mathcal{H}(\tau)\delta (\kappa - \kappa_0)$, where $1 \leq \kappa_0 \leq \kappa_\mathrm{max}$, as well as assuming the constant conductivity with $F=1$ and the maximal helicity density, at which $\mathcal{H}(\tau)=\mathcal{R}(\tau)/\kappa_0$ or $h(t)=2\rho_\mathrm{B}(t)/k_0$, we can recover the master equations in ref.~\cite{Dvornikov:2014uza}.

There is, however, a discrepancy between eq.~\eqref{system} and eq.~(16) in ref.~\cite{Dvornikov:2014uza}. It consists in the additional factor 2 in the $(\mathcal{M} + \mathcal{V})$ term in eq.~\eqref{system}. This factor appears since in ref.~\cite{Dvornikov:2014uza} we have relied on the incorrect eq.~(6) in ref~\cite{Boyarsky:2011uy}. The evolution of $h(k,t)$ is correctly described by eq.~\eqref{general} (see also refs.~\cite{Dvornikov:2013bca,Cam07}). Nevertheless, as we will see in sections~\ref{helicity-growth}-\ref{BGEN}, the main features of the magnetic field evolution, described in ref.~\cite{Dvornikov:2014uza}, remain unchanged.

Separating the magnetic diffusion factor $A_\mathrm{diff}(\tau)=\exp \left(-\kappa^2\int_{\tau_0}^{\tau}F(\tau')d\tau' \right)$ in the first two lines in eq.~(\ref{system}), one can easily find the important relation between the magnetic energy density spectrum $\mathcal{R}(\kappa,\tau)$ and
the magnetic helicity density spectrum $\mathcal{H}(\kappa,\tau)$,
\begin{equation}\label{relation}
  \mathcal{H}(\kappa,\tau) =
  \sqrt{
  \left[
    \frac{\mathcal{R}(\kappa,\tau)}{\kappa}
  \right]^2 -
  \left[
    \frac{\mathcal{R}(\kappa,\tau_0)}{\kappa}
  \right]^2
  (1 - q^2)},
  \quad
  0\leq q\leq 1.
\end{equation}
The relation in eq.~\eqref{relation} results from the conservation law for the auxiliary functions $\mathcal{H}_1(\kappa,\tau)=\mathcal{H}(\kappa,\tau)/A_\mathrm{diff}(\tau)$ and $\mathcal{R}_1(\kappa,\tau)=\kappa^2\mathcal{R}(\kappa,\tau)/A_\mathrm{diff}(\tau)$,
\begin{equation}
  \frac{\rm{d}}{\rm{d}\tau}
  \left[
    \kappa^2\mathcal{H}_1^2(\kappa,\tau) - \mathcal{R}_1^2(\kappa,\tau)
  \right] = 0,
\end{equation}
which obey the simplified differential equations, $\partial_\tau \mathcal{H}_1(\kappa,\tau) =2F(\tau) (\mathcal{M}(\tau) + \mathcal{V})\mathcal{R}_1(\kappa,\tau)$ and $\partial_\tau \mathcal{R}_1(\kappa,\tau)=2\kappa^2F(\tau)(\mathcal{M}(\tau) + \mathcal{V})\mathcal{H}_1(\kappa,\tau)$, obtained from eq.~(\ref{system}).

At the initial time $\tau_0$ the relation in eq.~(\ref{relation}) takes the form:
\begin{equation}\label{initialhelicity}
  \mathcal{H}(\kappa,\tau_0) = q
  \left[
    \frac{\mathcal{R}(\kappa,\tau_0)}{\kappa}
  \right].
\end{equation}
Here the parametrization by the factor $q\leq 1$ corresponds to the the relation $h(k,t_0)=2q\rho_\mathrm{B}(k,t_0)/k$ in dimensional notations. Thus only the particular case $q=1$ gives the maximal helicity density used in ref.~\cite{Dvornikov:2014uza}.

\subsection{Initial conditions\label{init-condition}}

In this section we choose the initial conditions for the system in eq.~(\ref{system}) corresponding to a realistic NS.

The value of $\mathcal{R}(\kappa,\tau_0)=(\alpha_\mathrm{em}^2/k_\mathrm{min}\mu^2)\rho_\mathrm{B}(k,t_0)$ is given by the {\it continuous} initial magnetic energy density spectrum \cite{Moffat,ZelRuzSok90},
\begin{equation}\label{initial_energy}
\rho_\mathrm{B}(k,t_0)=C k^{2 + \nu_\mathrm{B}},
\quad
k_\mathrm{min}\leq k\leq k_\mathrm{max},
\end{equation}
where the minimal wave number corresponds to the largest scale $\Lambda_\mathrm{B}=R_\mathrm{NS}$ for an {\it internal} magnetic field within the NS core with the radius $R_\mathrm{NS}=10\thinspace\mathrm{km}$, $k_\mathrm{min}=R_\mathrm{NS}^{-1}=2\times 10^{-11}\thinspace\text{eV}$. The maximal wave number $k_\mathrm{max}$ is a free parameter corresponding to the minimal spatial scale for the magnetic field $\Lambda_\mathrm{B}^{(\mathrm{min})}=k_\mathrm{max}^{-1}$.
The normalization constant $C$ in eq.~(\ref{initial_energy}) results from eq.~\eqref{rhodef} and equals to
\begin{equation}\label{normalization}
  C =
  \frac{(3 + \nu_\mathrm{B})B_0^2}{2k_\mathrm{max}^{3 + \nu_\mathrm{B}}}
\end{equation}
for $k_\mathrm{min}\ll k_\mathrm{max}$.

In eq.~\eqref{initial_energy}, we choose the Kolmogorov's spectrum for the initial energy density with $\nu_\mathrm{B}= - 5/3$, while other models of continuous spectra are possible, e.g. the Kazantsev's spectrum with $\nu_\mathrm{B}= - 1/2$, or the white noise case $\nu_\mathrm{B}=0$.

Basing on eqs.~(\ref{initial_energy}) and~(\ref{normalization}) and using eq.~(\ref{initialhelicity}), we choose finally the following initial conditions in eq.~(\ref{system}) for dimensionless quantities:
\begin{eqnarray}\label{initial}
  && \mathcal{H}(\kappa, \tau_0) = q\mathcal{C}
  \left(
    \frac{\kappa}{\kappa_\mathrm{max}}
  \right)^{1 +\nu_\mathrm{B}},
  \quad
  1\leq \kappa\leq \kappa_\mathrm{max} =
  \frac{k_\mathrm{max}}{k_\mathrm{min}}\gg 1,
  \nonumber
  \\
  && \mathcal{R}(\kappa, \tau_0) = \mathcal{C}
  \left(
    \frac{\kappa}{\kappa_\mathrm{max}}
  \right)^{2 + \nu_\mathrm{B}},
  \quad
  \mathcal{C} =
  \left(
    \frac{\alpha_\mathrm{em}^2
    (3 + \nu_\mathrm{B})B_0^2}{2\mu^2k_\mathrm{max}^2}
  \right),
  \nonumber
  \\
  && \mathcal{M}(\tau_0) =
  \frac{\alpha_\mathrm{em}}{\pi k_{\min}}\mu_5(t_0) =
  1.2\times 10^{14}.
\end{eqnarray}
Here in the last line $\mu_5(t_0)=1\thinspace{\rm MeV}\ll \mu$ was used as an arbitrary value at the beginning of the chiral imbalance production through an electroweak mechanism at the earlier stages $t < t_0$ of the NS evolution.


%

\subsection{Growth of electric conductivity for a cooling non-superfluid NS\label{t0}}

In this section we study the influence of the NS cooling on the time dependence of the conductivity and thus on the growth of the magnetic field.

We remind that, in our problem, $\sigma_\mathrm{cond}$ is the electric conductivity in a degenerate electron-proton plasma consisting of ultrarelativistic degenerate electrons and non-relativistic degenerate protons. The effects of both $ee$-collisions and the scattering of electrons by a neutron magnetic moment are minor for the electric conductivity in plasma, $\sigma_\mathrm{cond}=\omega_p^2/\nu_\mathrm{coll}$, given mostly by $ep$-collisions \cite{Kelly}. Here $\omega_p=\mu\sqrt{4\alpha_\mathrm{em}/3\pi}$ is the plasma frequency in a degenerate ultrarelativistic electron gas with the density $n_e=\mu^3/3\pi^2$. Note that in a degenerate electron gas both $\nu_\mathrm{coll}$ and $\sigma_\mathrm{cond}$ depend on the temperature $T$ since $\nu_\mathrm{coll}\sim T^2$ . This is due to the Pauli principle when all electron states with the electron momenta $0\leq p_e\leq \mu$ are busy, i.e. the $ep$-scattering is impossible at $T=0$.

For the constant electron abundance $Y_e=n_e/n_\mathrm{B}\approx \text{const}$ the electric conductivity rises with the cooling of a NS core~\cite{Kelly}\footnote{The dependence of $\sigma_\mathrm{cond}$ in eq.~(\ref{conductivity}) on the electron abundance, $Y_e=n_e/n_\mathrm{B}$, where $n_\mathrm{B}=n_n + n_p\approx n_n$, is more slower than that on the temperature. For a cooling NS at the second stage the conventional value $Y_e\simeq 0.05$ corresponds to $n_e=9\times 10^{36}\thinspace\text{cm}^{-3}$},
\begin{equation}\label{conductivity}
  \sigma_\mathrm{cond}(t)=\sigma_0
  \left[
    \frac{T(t_0)}{T(t)}
  \right]^2,
  \quad
  \sigma_0=10^7
  \left(
    \frac{n_e}{10^{36}\thinspace\text{cm}^{-3}}
  \right)^{3/2}\thinspace{\rm MeV}\approx \text{const},
\end{equation}
due to the negative derivative $\mathrm{d}T/\mathrm{d}t <0$ given by the neutrino (antineutrino) emission~\cite{Pethick,Yakovlev}:
\begin{equation}\label{law}
 \frac{\mathrm{d}T(t)}{\mathrm{d}t}= -\frac{T(t)}{(n_\mathrm{T}-2)t},
 \end{equation}
where the index $n_\mathrm{T}$ equals to 6 or 8 depending on the neutrino emission channel\footnote{For the redshifted spatially constant internal temperature $T(t)=T(r,t)\exp [\Phi (r)]$, where $T(r,t)$ is the local internal temperature, $r$ is the radial coordinate and $\Phi(r)$ is the metric function that determines the gravitational redshift, we do not take into account in eq.~(\ref{law}) the constant
factor $\exp [\Phi(r)]\approx \exp [\Phi (R_\mathrm{NS})]$ and consider the uniform internal temperature $T(r,t)\approx T_b$ where $T_b$ is temperature at the bottom of a thin envelope of NS related to the observable surface temperature by $T_s\propto \sqrt{T_b}$; cf. ref.~\cite{Yakovlev}.}.

At early times, direct Urca processes, $n\to p + e^- + \bar{\nu}_e$ and $p + e^-\to n + \nu_e$, prevail  with the neutrino emissivity $L_\mathrm{dU}\sim 10^{34}(T/10^8\thinspace\text{K})^6 \thinspace \text{erg}\cdot\text{s}^{-1}$. Thus, at this stage the index $n_\mathrm{T}=6$ should be substituted in eq.~(\ref{law}).
These equilibrium processes start just after a SN burst at first seconds, when the electron abundance is large enough $Y_e\sim 0.4$. However, then the electron abundance $Y_e$ becomes less due to the deleptonization within the NS core as the SN remnant, and below $Y_e< 1/9$ the direct Urca processes turn out to be suppressed because degenerate fermions are unable to conserve   momentum while remaining on their Fermi surfaces. However, the presence of a third body, a nucleon $N$, in the modified Urca processes, $n + N\to p + e^- + N + \bar{\nu}_e$ and $p + N + e^-\to n + N + \nu_e$, allows to conserve momentum, and the neutrino (antineutrino) emissivity
$L_\mathrm{mU}\sim 10^{30}(T/10^8\thinspace\text{K})^8\thinspace \text{erg}\cdot\text{s}^{-1}$, corresponding to the index $n_\mathrm{T}=8$, dominates at $Y_e< 1/9$. Thus we will adopt here $n_\mathrm{T}=8$ in eq.~(\ref{law}) assuming also the conventional constant electron (proton) abundance $Y_e\sim 0.05$.

From eqs.~(\ref{conductivity}) and~(\ref{law}) one can find the factor $F=\sigma_0/\sigma_\mathrm{cond} (t)$ in eq.~(\ref{system}),
\begin{equation}\label{F}
  F=
  \left[
    \frac{T(t)}{T(t_0)}
  \right]^2 =
  \left(
    \frac{t}{t_0}
  \right)^{-2/(n_\mathrm{T}-2)},
\end{equation}
or $F(\tau)= (\tau/\tau_0)^{-1/3}$ for $n_\mathrm{T}=8$.
Here we consider the cooling problem for a {\it thermally relaxed non-superfluid NS core} at the neutrino cooling stage.
Before this time interval, during the first stage which lasts from $\sim 10\thinspace\text{yr}$ to a few centuries, a newly born star is thermally non-relaxed. We
put here $t_0=100\thinspace\text{yr}$ ($T_0=10^8\thinspace\text{K}$) as the initial time (temperature) for our evolution eq.~(\ref{system}) and consider here the second stage which lasts for $(10^5 - 10^6)\thinspace\text{yr}$, $t_0\leq t\leq 10^6\thinspace\text{yr}$, when the core is thermally decoupled from the crust and it cools via the neutrino emission. During the third (final) stage, $t> 10^6\thinspace\text{yr}$, a thermally relaxed NS cools via the surface emission of thermal photons. Thus, the time $t_0=100\thinspace\text{yr}<t< 10^6\thinspace\text{yr}$ is appropriate in our numerical calculations where we use the cooling law in eq.~(\ref{law}).

	
\subsection{Magnetic helicity growth driven by weak $eN$ interaction\label{helicity-growth}}

In this section we study the evolution of the magnetic helicity based on the numerical solution of the system in eq.~\eqref{system}.

\begin{figure}
  \centering
  \subfigure[]
  {\label{1a}
  \includegraphics[scale=.36]{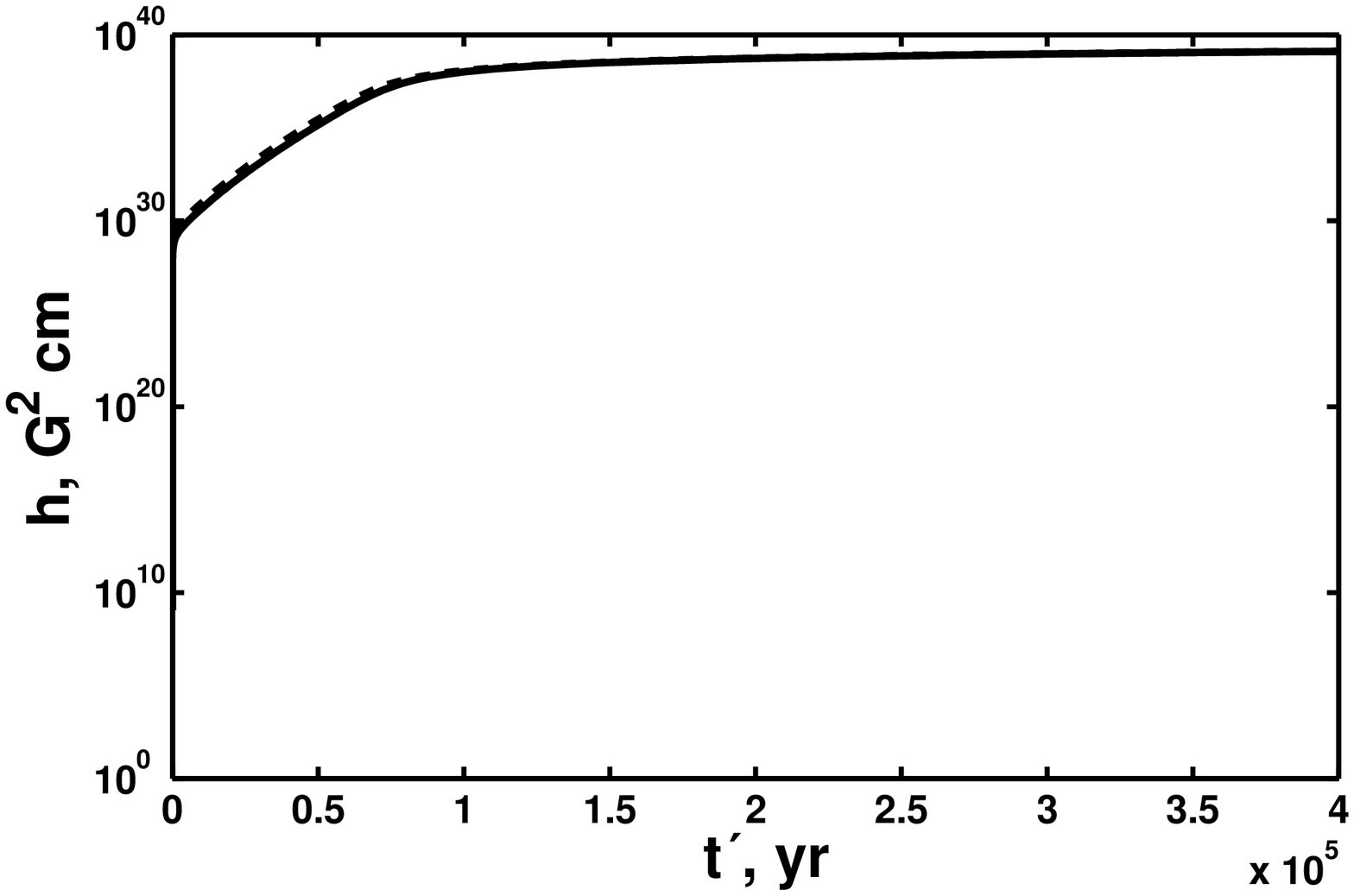}}
  \hskip-.7cm
  \subfigure[]
  {\label{1b}
  \includegraphics[scale=.36]{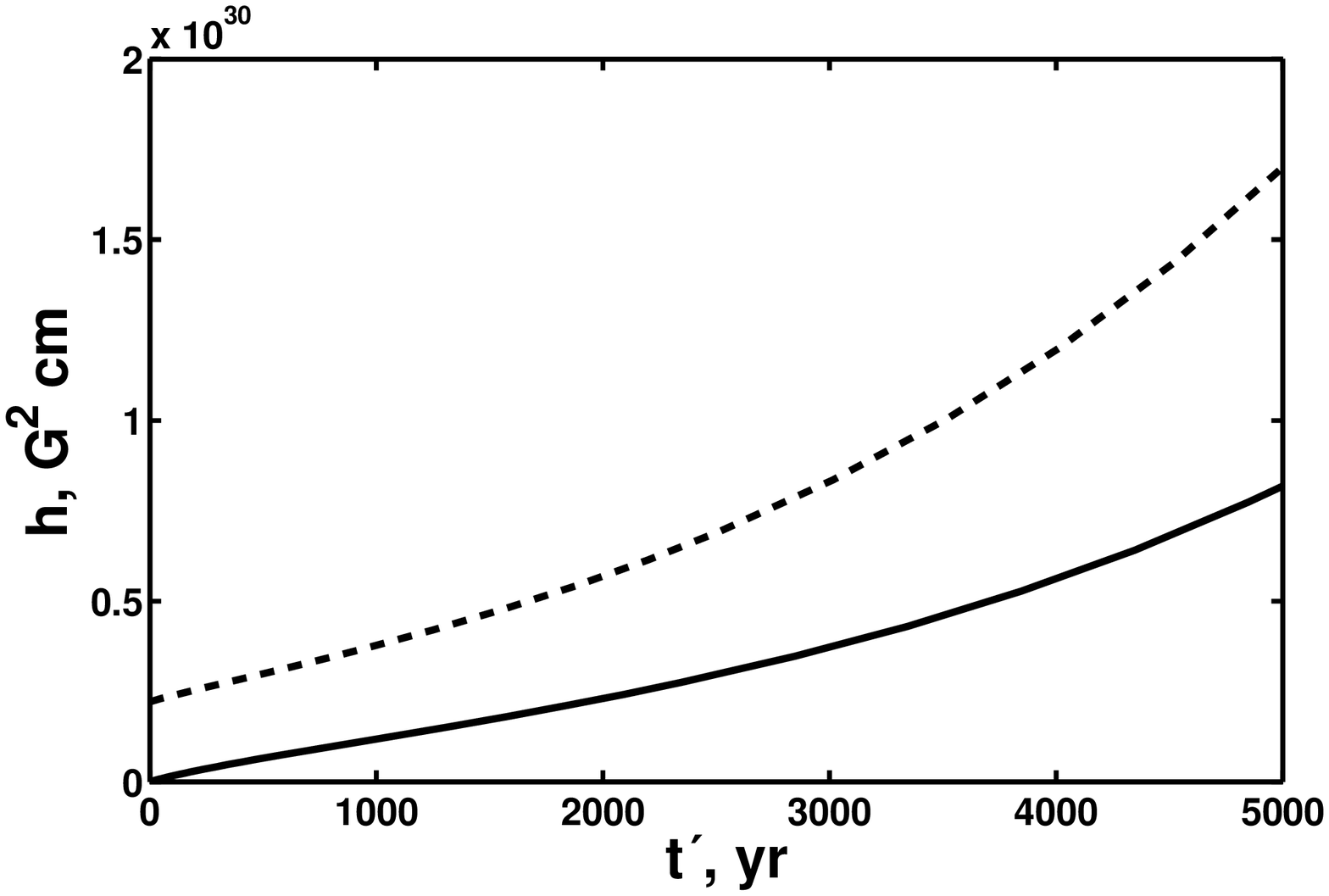}}
  \\
  \subfigure[]
  {\label{1c}
  \includegraphics[scale=.36]{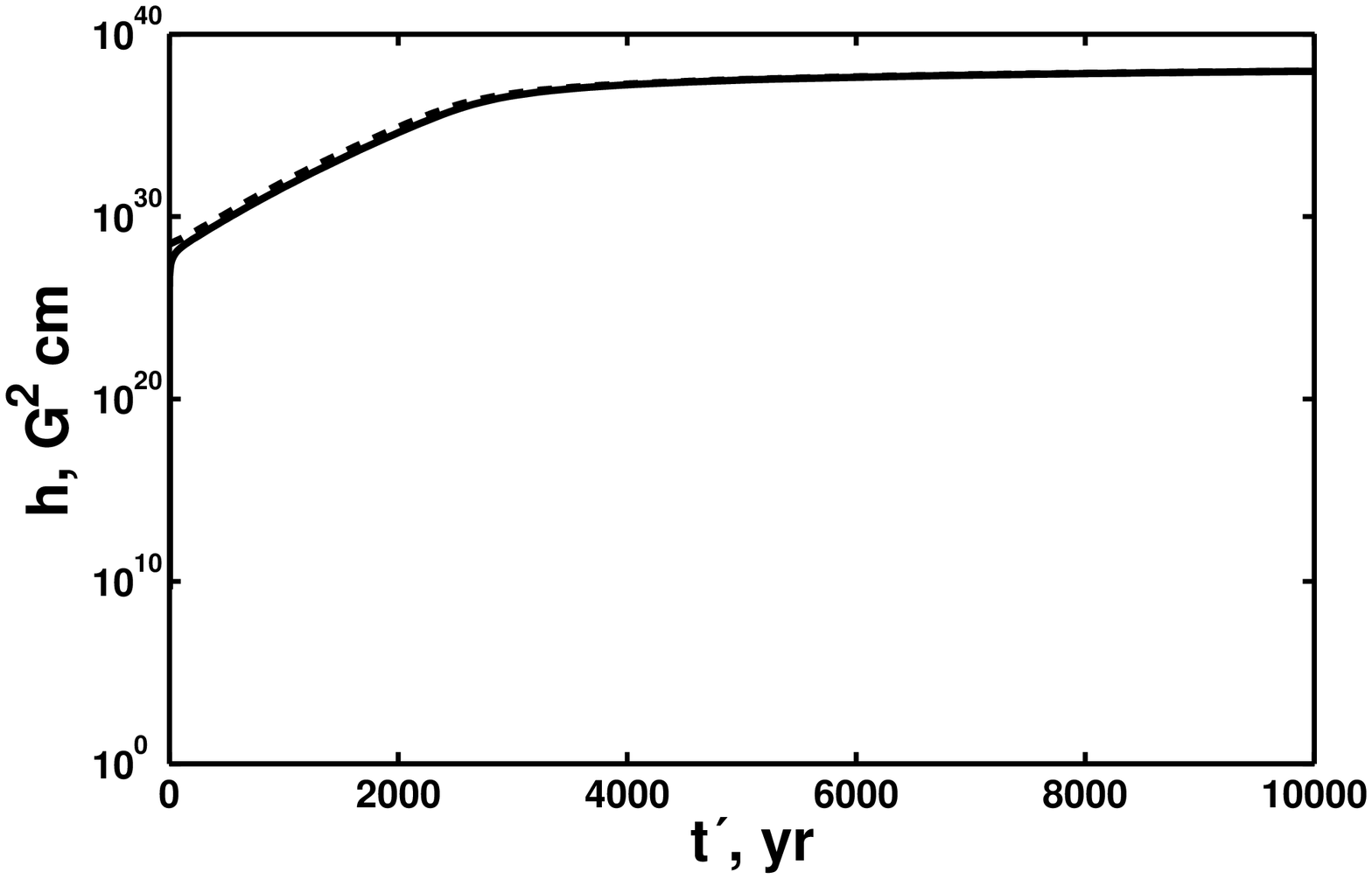}}
  \hskip-.7cm
  \subfigure[]
  {\label{1d}
  \includegraphics[scale=.36]{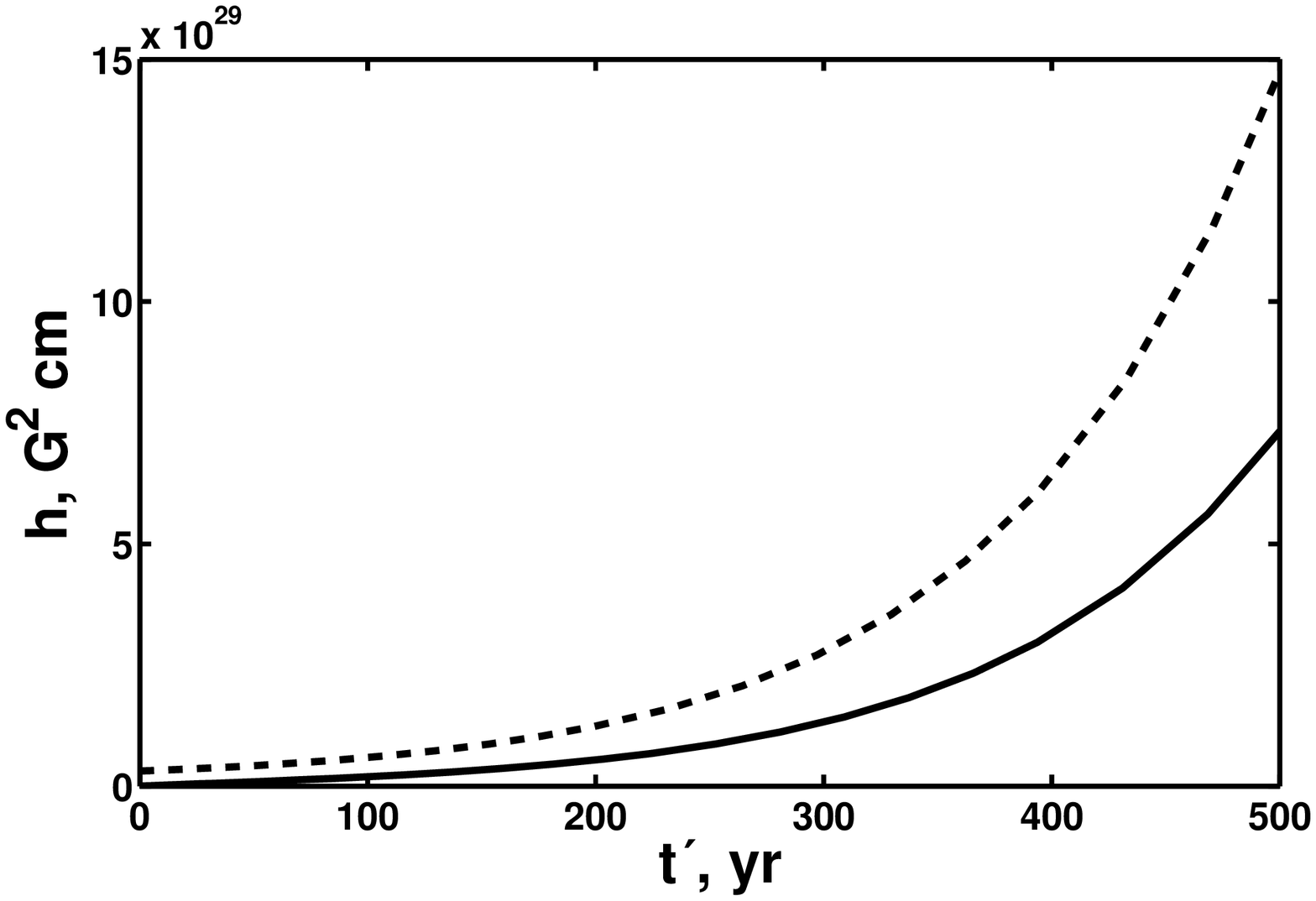}}
    \caption{The dimensional (in standard units ${\rm G^2cm}$) helicity density $h(t) = (2\mu^2k_\mathrm{max}/\alpha_\mathrm{em}^2) \int_1^{\kappa_\mathrm{max}}\mathrm{d}\kappa\mathcal{H}(\kappa,\tau)$ versus time $t' = t - t_0$ for the initially non-helical field, $q=0$ (solid lines), and the maximum helical field, $q=1$ (dashed lines). Here $k_\mathrm{min}=R_\mathrm{NS}^{-1}=2\times 10^{-11}\thinspace\text{eV}$ and $t_0=100\thinspace\text{yr}$.
    Panel (a) corresponds to $k_\mathrm{max}=2\times 10^{-10}\thinspace\text{eV}$, or $\Lambda_\mathrm{B}^{(\mathrm{min})}=1\thinspace\mathrm{km}$. The helicity evolution is shown for $t_0 < t < 4 \times 10^5\thinspace\text{yr}$. Panel (b) is the same as the panel (a) but for earlier times $t_0 < t < 5 \times 10^3\thinspace\text{yr}$.
    Panel (c) corresponds to $k_\mathrm{max}=2\times 10^{-9}\thinspace\text{eV}$, or $\Lambda_\mathrm{B}^{(\mathrm{min})}=100\thinspace\text{m}\ll R_\mathrm{NS}$, for $t_0 < t < 10^4 \thinspace \text{yr}$.
    Panel (d) is the same as the panel (c) but for earlier times $t_0 < t < 6\times 10^2 \thinspace\text{yr}$.
  \label{fig:h}}
\end{figure}

In figure~\ref{fig:h} we show the corresponding growth of the helicity density $h(t)=\smallint_{k_\mathrm{min}}^{k_\mathrm{max}} \mathrm{d}kh(k,t)$ in the case of the Kolmogorov's spectrum with $\nu_\mathrm{B}=-5/3$. Figures~\ref{1a} and~\ref{1b} correspond to $k_\mathrm{max}=2\times 10^{-10}\thinspace\text{eV}$ or $\Lambda_\mathrm{B}=1\thinspace\mathrm{km}$, and figures~\ref{1c} and~\ref{1d} to $k_\mathrm{max}=2\times 10^{-9}\thinspace\text{eV}$ or $\Lambda_\mathrm{B}^{(\mathrm{min})}=100\thinspace\text{m}\ll R_\mathrm{NS}$.  To support the helicity growth, i.e. to have $\mathrm{d}h/\mathrm{d}t>0$, the maximal wave number $k_\mathrm{max}$, that is equivalent to the smallest scale $\Lambda_\mathrm{B}=(k_\mathrm{max})^{-1}$, should obey the inequality\footnote{Using eq.~(\ref{general}) one can show that the helicity growth is possible just at the initial time $t_0$ if
$$
  \left.\frac{\mathrm{d}h(k,t)}{\mathrm{d}t}\right|_{t=t_0} =
  \frac{2\rho_\mathrm{B}(k, t_0)}{\sigma_0}\left[- 2qk + \Pi(t_0)\right]>0.
$$}:
\begin{equation}\label{inequality}
  kq < \frac{\Pi(t_0)}{2},
  \quad
  \Pi(t_0)=\frac{2\alpha_\mathrm{em}}{\pi}(\mu_5(t_0) + V_5).
\end{equation}
One can easily see the condition in eq.~\eqref{inequality} is fulfilled for any chiral imbalance $\mu_5$ since $V_5=6\thinspace\text{eV} = \text{const} \gg kq$ for any $k\leq k_\mathrm{max}$.

The most interesting issue here is the growth of the helicity density
$h(t)$ for the {\it initially non-helical field}, $q=0$.
It is remarkable that the difference of the helicity densities for the initially non-helical magnetic field, $q=0$, shown by the solid lines, and the maximum helical field, $q=1$ , shown by the dashed lines, vanishes at $t \gg t_0$. In other words, an initial non-helical field tends to be maximum helical irrespective to the initial condition in eq.~(\ref{initial}); cf. figures~\ref{1a} and~\ref{1c}. On the contrary, at earlier times $t \gtrsim t_0$ such a difference is significant; cf. figures~\ref{1b} and~\ref{1d}. In all cases we start from $t_0=100\thinspace\text{yr}$ corresponding to the initial moment of the thermally relaxed non-superfluid NS core (see the comments in section~\ref{t0}).

While the minimal wave number $k_\mathrm{min}=R_\mathrm{NS}^{-1}$ is fixed, we can not choose rather big values of $k_\mathrm{max}$ or the ratio $\kappa_\mathrm{max}=k_\mathrm{max}/k_\mathrm{min}\gg 1$. Otherwise,
the minimal scale of magnetic field $\Lambda_\mathrm{B}^{(\mathrm{min})}=k_\mathrm{max}^{-1}$ could be  comparable with a small scale of the fluid velocity
$\lambda_v$ we neglected here, when we consider large-scale magnetic fields and put $\lambda_v\ll k^{-1}$, in order to avoid the involvement of the Navier-Stokes equation in our simplified model.

Considering smaller scales $\Lambda_\mathrm{B}=k^{-1}$, compared to those shown in lower panels in figure~\ref{fig:h}, we should also discuss earlier initial times $t_0$, when the stage of a thermally relaxed non-superfluid NS core have not been started yet. In other words, one expects that for a big $k_\mathrm{max}$ the medium should be turbulent. This fact is not surprising because for small scales, i.e. when $k$ is big, the evolution of $h(t)$ and $\rho_\mathrm{B}(t)$  proceeds faster since both characteristics, like the helicity density and the magnetic energy density, are proportional to the running wave number $k$, $h(k,t)\sim A_k(t) B_k(t)\simeq k A_k^2$ and $\rho_\mathrm{B} (k,t)\simeq k^2A_k^2$, involved in continuous spectra. The example of such an accelerated growth can be seen in figures~\ref{1c} and~\ref{1d}, where $k_\mathrm{max}=2\times 10^{-9}\thinspace\text{eV}$, or $\Lambda_\mathrm{B}^{(\mathrm{min})}=100\thinspace\text{m}\ll R_\mathrm{NS}$, is chosen.
\subsection{Change of the sign for CME\label{SINGCME}}

In this section we discuss some peculiarities in the  evolution of the chiral imbalance $\mu_5$.

\begin{figure}
  \centering
  \subfigure[]
  {\label{2a}
  \includegraphics[scale=.36]{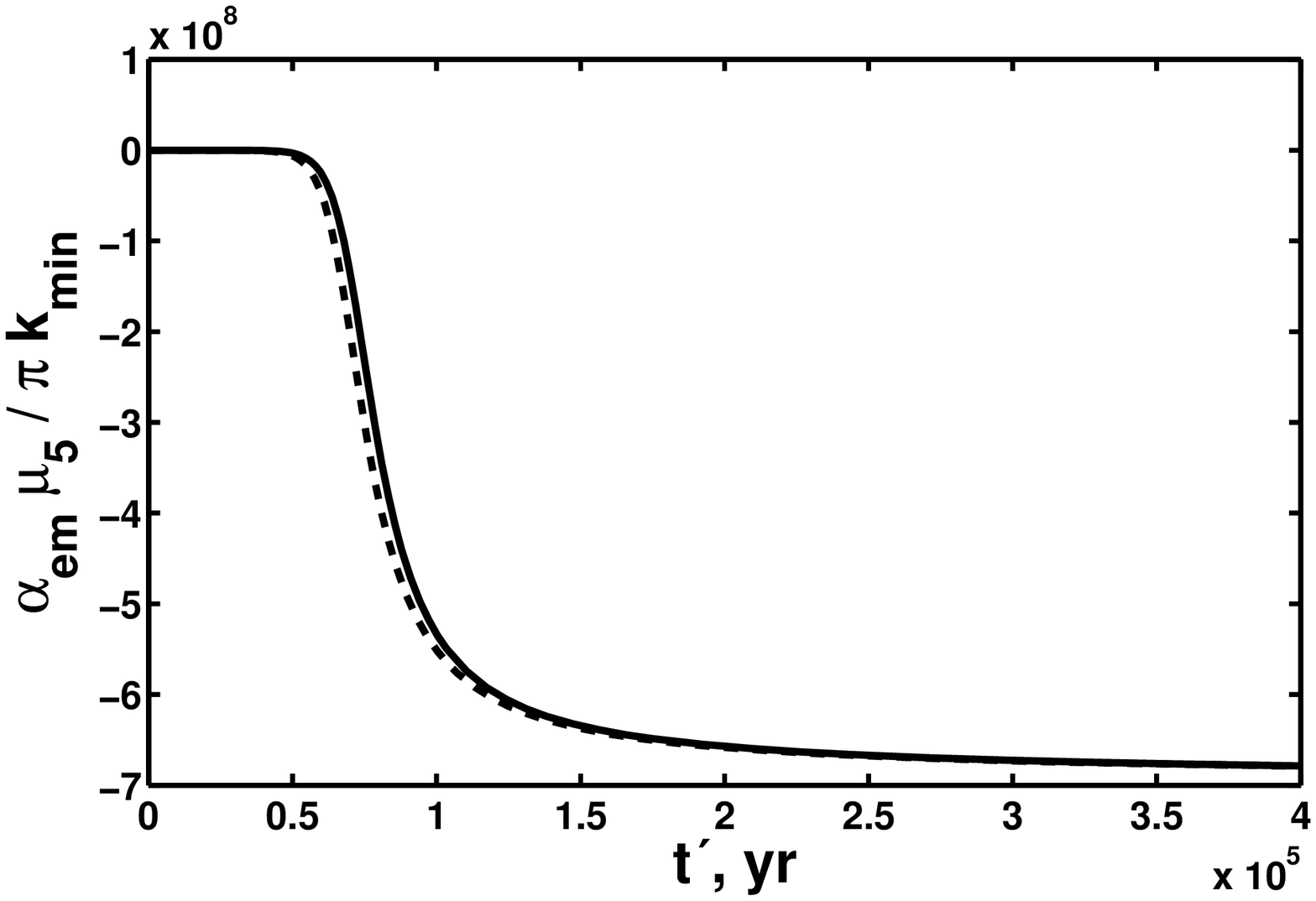}}
  \hskip-.7cm
  \subfigure[]
  {\label{2b}
  \includegraphics[scale=.36]{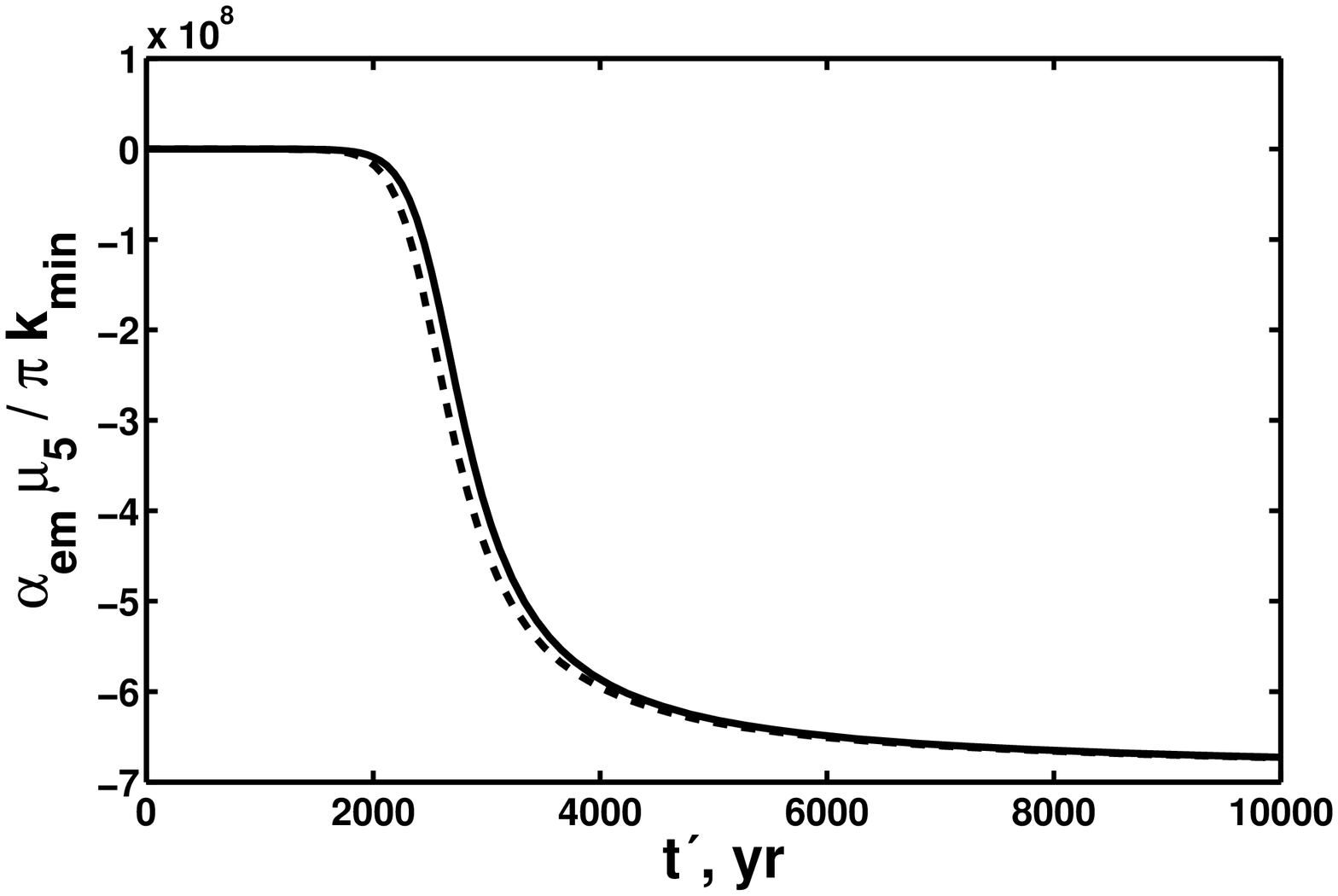}}
    \caption{The dimensionless chiral imbalance $\mathcal{M}$ versus time $t' = t - t_0$ for the initially non-helical field, $q=0$ (solid line), and the maximum helical field, $q=1$ (dashed line). We choose $t_0=100\thinspace\text{yr}$ and $k_\mathrm{min}=R_\mathrm{NS}^{-1}=2\times 10^{-11}\thinspace\text{eV}$.
    Panel (a) corresponds to $k_\mathrm{max}=2\times 10^{-10}\thinspace\text{eV}$, or $\Lambda_\mathrm{B}^{(\mathrm{min})}=1\thinspace\mathrm{km}$, and $t_0 < t < 4 \times 10^5\thinspace\text{yr}$.
    Panel (b) is built for $k_\mathrm{max}=2\times 10^{-9}\thinspace\text{eV}$, or $\Lambda_\mathrm{B}^{(\mathrm{min})}=100\thinspace\text{m}$, and $t_0 < t < 10^4\thinspace\text{yr}$.
  \label{fig:mu5}}
\end{figure}

In figure~\ref{fig:mu5} we show the evolution of the CME parameter $\mathcal{M}\sim \mu_5$ at different time scales based on the numerical solution of the system in eq.~\eqref{system}. In section~\ref{init-condition} we have already commented on an initial positive $\mu_5(t_0)\sim O(\mathrm{MeV})\ll \mu=125\thinspace\text{MeV}$ that vanishes fast during $t\sim 10^{-12}\thinspace\text{s}\ll t_0$ due to the huge chirality flip rate $\mathcal{G}\sim 10^{30}$~\cite{Dvornikov:2014uza}. This fast relaxation of $\mu_5$ to zero is not shown in figure~\ref{fig:mu5}. The example of the  attenuation of $\mu_5 \to 0$ is given in ref.~\cite{Dvornikov:2014uza}. The curves in figure~\ref{fig:mu5} start from the plateau $\mathcal{M}\approx 0$. In figure~\ref{2b} one can see that this plateau lasts for $\sim 2000\thinspace\text{yr}\gg t_0$ for the maximal $k_\mathrm{max}=2\times 10^{-9}\thinspace\text{eV}$. Then the CME parameter $\mu_5(t)$ changes sign, and for the same $k_\mathrm{max}$ reaches the absolute value of the weak interaction parameter $V_5$, $|\mathcal{M}| \to \mathcal{V}=7\times 10^8$, or $|\mu_5| \to V_5=6\thinspace\text{eV}$. Nevertheless, as seen in figure~\ref{fig:mu5}, the sum $\mathcal{M} + \mathcal{V}$ remains positive, $\mathcal{M} + \mathcal{V}>0$, that provides the magnetic helicity growth, ${\rm d}\mathcal{H}/{\rm d}\tau > 0$, and the magnetic field strength itself, ${\partial}\mathcal{R}/{\partial}\tau > 0$, as it results from eq.~(\ref{system}).

The change of the sign for the CME parameter $\mu_5$ at a certain moment of time, $\mu_5>0\to \mu_5<0$, is explained by the huge negative back reaction from the growing magnetic field, when the second term in eq.~(\ref{kinetics2}), $\sim \rho_\mathrm{B}(t)$, corresponding to the second negative term in the last line in eq.~(\ref{system}) for ${\rm d}\mathcal{M}/{\rm d}\tau$, becomes much greater than the chirality flip term, $\sim \Gamma_f$. For the initial magnetic field, $B_0=10^{12}\thinspace\text{G}$, or at earlier times $t< 2000\thinspace\text{yr}$, the situation is reversal: the back reaction from a moderate magnetic field is negligible and the chirality flip term is the main one, which reduces the initial positive $\mu_5>0$ to zero, $\mathcal{M}\to 0$.

\section{Generation of magnetic fields in magnetars driven by weak $eN$ interactions\label{BGEN}}

The previous attempts to explain the growth of a seed field $B_0\sim 10^{12}\thinspace\text{G}$ in a NS core up to the strongest  $B=(10^{15}-10^{16})\thinspace\text{G}$ observable in magnetars \cite{Mer08},  which were based, e.g., on CME with $\mu_5\neq 0$ in refs.~\cite{Yamamoto2,Yamamoto1}, failed because of the underestimated chirality flip $\sim \Gamma_f$, see comments on this issue in refs.~\cite{Dvornikov:2014uza,grabow,Kha15,Miransky:2015ava}.

\begin{figure}
  \centering
  \subfigure[]
  {\label{3a}
  \includegraphics[scale=.36]{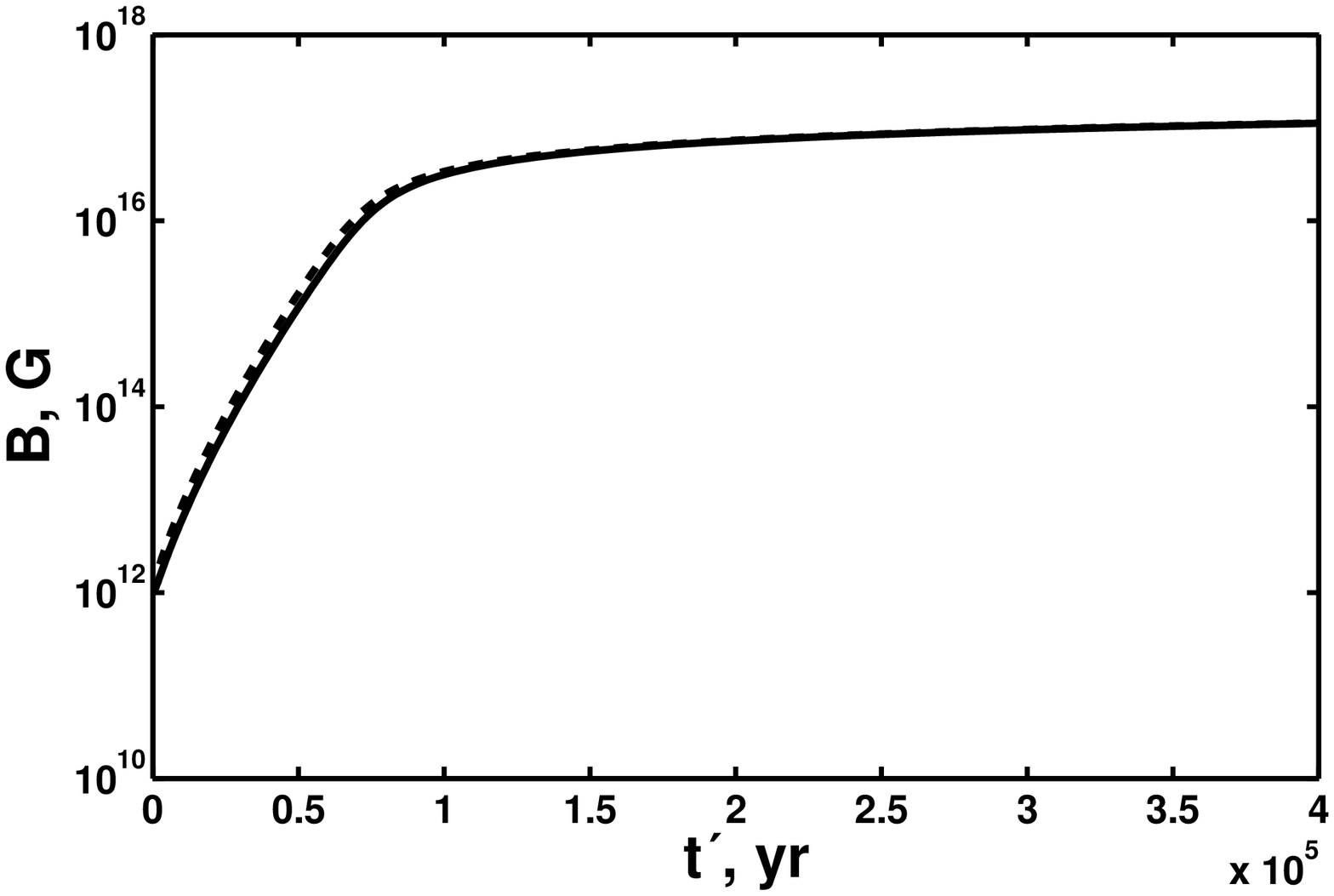}}
  \hskip-.7cm
  \subfigure[]
  {\label{3b}
  \includegraphics[scale=.36]{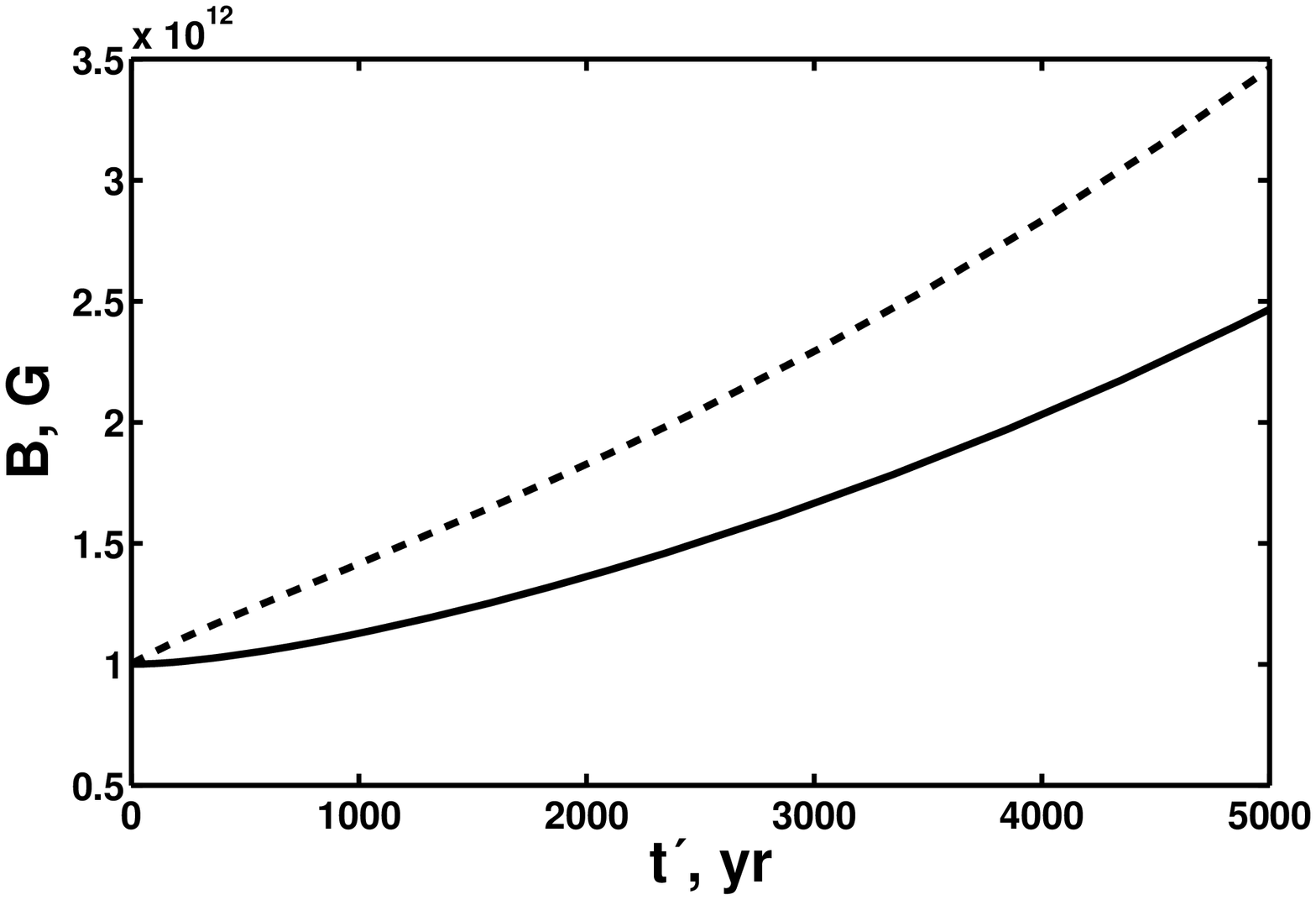}}
  \\
  \subfigure[]
  {\label{3c}
  \includegraphics[scale=.36]{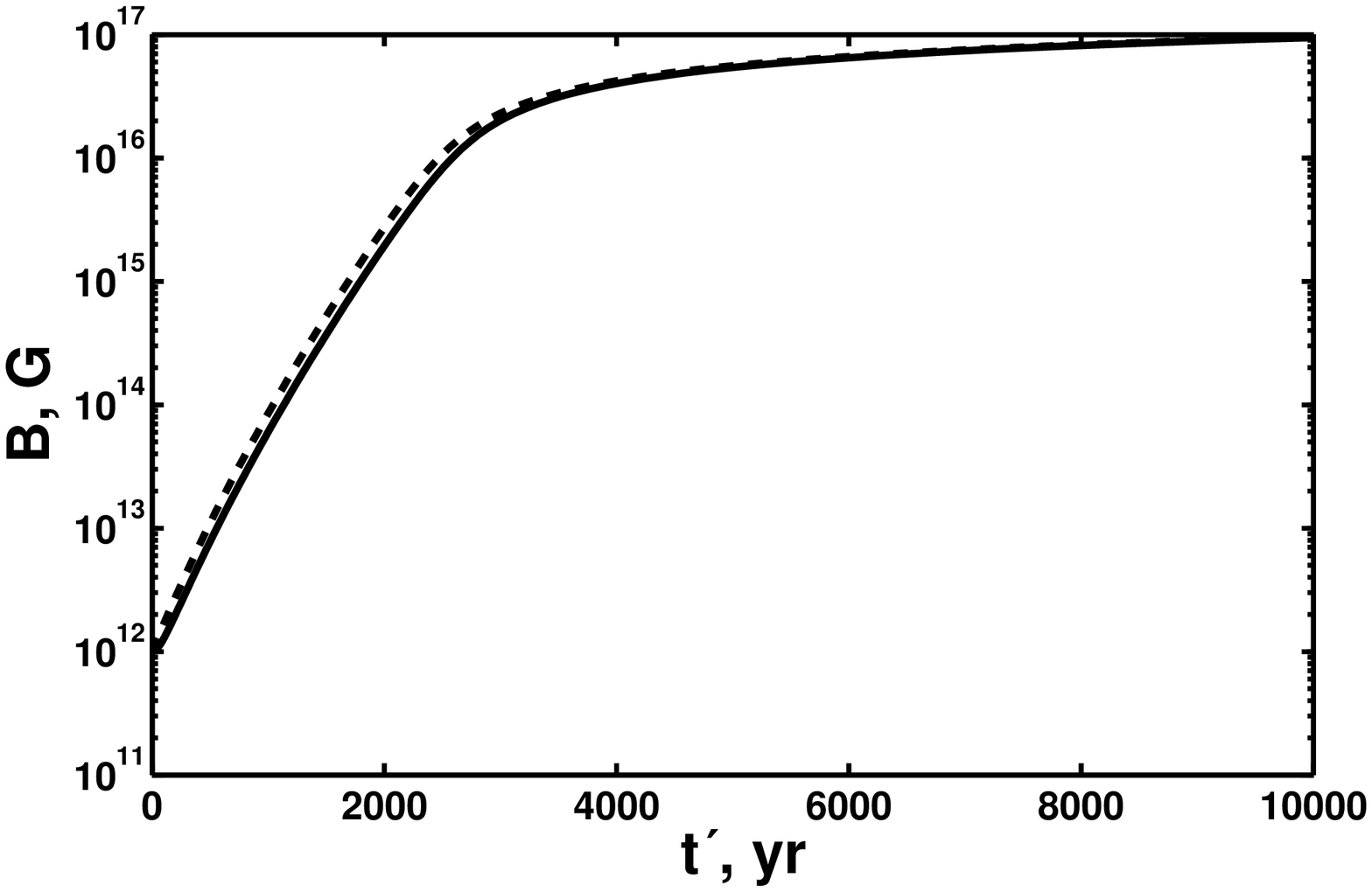}}
  \hskip-.7cm
  \subfigure[]
  {\label{3d}
  \includegraphics[scale=.36]{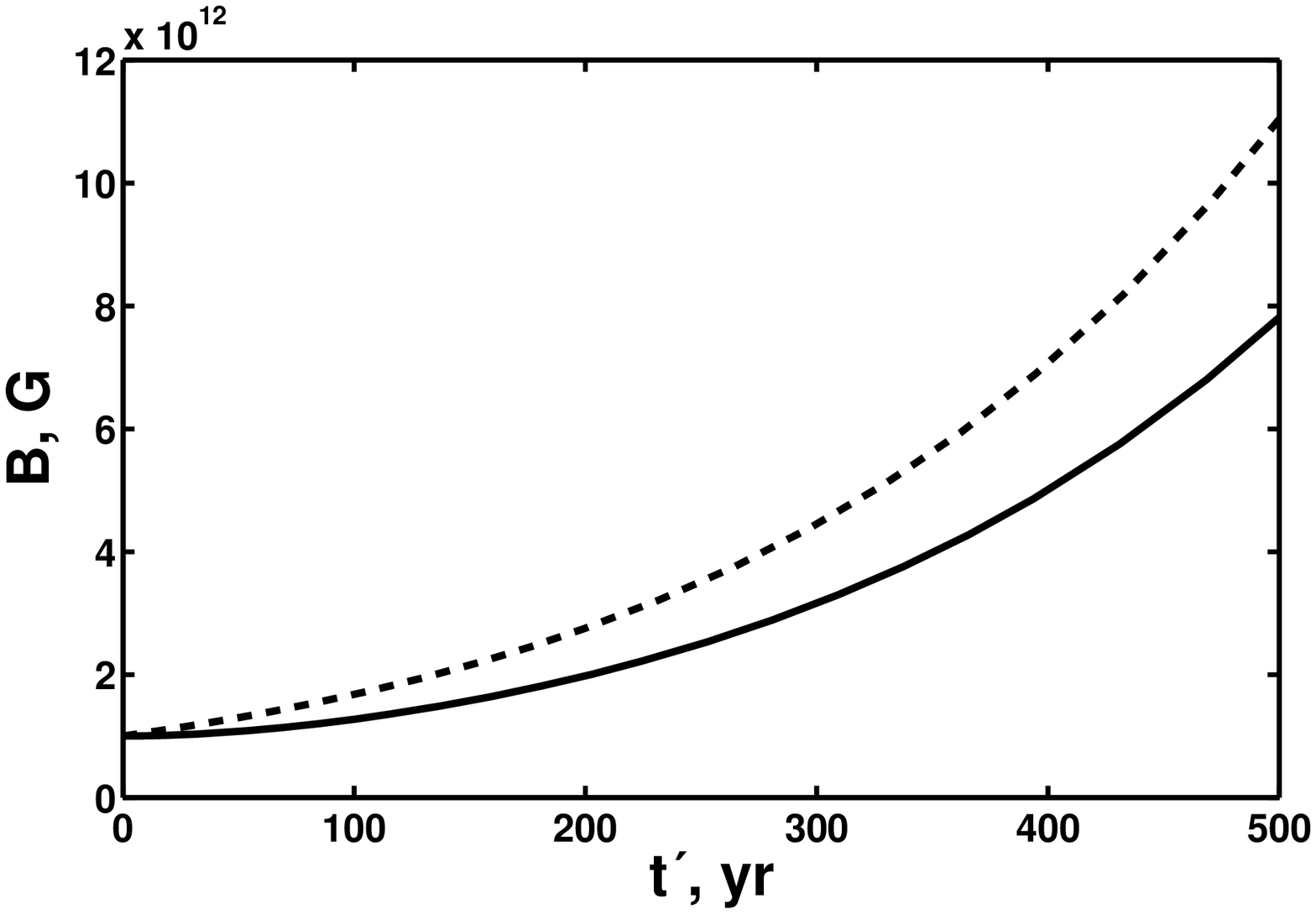}}
    \caption{The dimensional (in Gauss) magnetic field in NS, $B(t)=\sqrt{2\int dk\rho_\mathrm{B}(k,t)}$, versus time $t' = t - t_0$ for the initially non-helical field, $q=0$ (solid lines), and the maximum helical field, $q=1$ (dashed lines). Here $k_\mathrm{min}=R_\mathrm{NS}^{-1}=2\times 10^{-11}\thinspace\text{eV}$ and $t_0=100\thinspace\text{yr}$. Analogously to figures~\ref{fig:h} and~\ref{fig:mu5}, the Kolmogorov's initial spectrum $\rho_\mathrm{B}(k,t_0)$, given by eq.~(\ref{initial_energy}) with $\nu_\mathrm{B}=-5/3$, is assumed.
    Panel (a) corresponds to $k_\mathrm{max}=2\times 10^{-10}\thinspace\text{eV}$, or $\Lambda_\mathrm{B}^{(\mathrm{min})}=1\thinspace\mathrm{km}$. The magnetic field evolution is shown for $t_0 < t < 4 \times 10^5\thinspace\text{yr}$. Panel (b) is the same as panel (a) but at earlier times $t_0 < t < 5 \times 10^3\thinspace\text{yr}$. Panel (c) corresponds to $k_\mathrm{max}=2\times 10^{-9}\thinspace\text{eV}$, or $\Lambda_\mathrm{B}^{(\mathrm{min})}=100\thinspace\text{m}\ll R_\mathrm{NS}$, and times $t_0 < t < 10^4 \thinspace\text{yr}$.
    Panel (d) is the same as the panel (c) but for earlier times $t_0 < t < 6 \times 10^2 \thinspace\text{yr}$.
  \label{fig:B}}
\end{figure}

In figure~\ref{fig:B}, we show that the magnetic field instability caused by the current in eq.~(\ref{current}) leads to the growth of a seed field $B_0$ by about five orders of magnitude. The resulting magnetic field grows up to $B\simeq 10^{17}\thinspace\text{G}$ for $B_0=10^{12}\thinspace\text{G}$. The magnetic field growth happens in the interval $(10^3 - 10^5)\thinspace\text{yr}$, depending on the minimal scale $\Lambda_\mathrm{B}^{(\mathrm{min})}=k_\mathrm{max}^{-1}$ of the continuous spectrum in eq.~(\ref{initial_energy}). We revealed above the leading role of the weak interaction term $V_5$ for the magnetic field instability providing the growth of the magnetic field strength, $\partial \mathcal{R}/\partial \tau >0$.

Now it should be noted that our model (see also ref.~\cite{Dvornikov:2014uza}) does not require any special initial conditions, like an extremely strong magnetic field in a protostar in ref.~\cite{VinKui06}, or a fast rotation in ref.~\cite{Duncan}, or a significant initial chiral imbalance in refs.~\cite{Yamamoto2,Yamamoto1,Charbonneau:2009ax}, to generate strong magnetic fields in magnetars. Thus, using our results, we can predict a rather high magnetars abundance. This fact is in agreement with the findings of ref.~\cite{Woo08} that the number of magnetars in our universe should be comparable with that of ordinary pulsars having moderate magnetic fields $\sim B_0 = 10^{12}\thinspace\text{G}$.

\section{Discussion\label{CONCL}}

There are two anomalies in SM that give two corresponding terms in the electric current in eq.~(\ref{current}), ${\bf J}={\bf J}_\mathrm{CME} + {\bf J}_\mathrm{CS}$. The former one is the Adler anomaly that leads to the well-known current in QCD plasma, ${\bf J}_\mathrm{CME}= (2\alpha_\mathrm{em}/\pi)\mu_5{\bf B}$, see, e.g., ref.~\cite{Kharzeev} or appendix~A in ref.~\cite{DS1}. The latter anomaly leading to the Chern-Simons (CS) term ${\bf J}_\mathrm{CS}\sim G_\mathrm{F}$ arises due to the polarization effect when  accounting for the parity violation in SM; cf. ref.~\cite{Semikoz:2011tm}.
The corresponding CS term in the SM Lagrangian $L_\mathrm{CS}=\Pi_2({\bf A}\cdot{\bf B})={\bf A}\cdot{\bf J}_\mathrm{CS}$ is given by the static limit of the parity violation term in the photon polarization operator $\Pi_{ij}\sim e_{ijn}k_n\Pi_2(\omega = 0,|\mathbf{k}|)$~\cite{BoyRucSha12}. It arises  as the second term in eq.~(\ref{helicity}) for large scale magnetic fields, $|\mathbf{k}| \to 0$, $\Pi_2(0,0)=(2\alpha_\mathrm{em}/\pi)V_5\sim G_\mathrm{F}$. In the present work we did not apply the Matsubara technique  to derive such $\Pi_2$, as it was made in our previous works~\cite{Dvornikov:2013bca,Dvornikov2} for the electroweak $\nu e$-  and $ee$-interactions. Instead we adopted a more transparent method to derive ${\bf J}_\mathrm{CS}=\Pi_2{\bf B}$ for massless electrons basing on the exact solution of the Dirac equation, accounting for the electroweak $eN$ interaction under the influence of an external magnetic field (see ref.~\cite{Dvornikov:2014uza} and Appendix~\ref{CURRDER}). This approach includes automatically the CME term ${\bf J}_\mathrm{CME}$ in the current. Thus both terms, $\sim \mu_5$ and $\sim V_5$, are present in the kinetic equations in eq.~(\ref{system}). The anomalous current ${\bf J}$, being additive to the ohmic current enters Maxwell equation, results in the nonzero magnetic helicity parameter $\alpha=\Pi/\sigma_\mathrm{cond}$, which governs the Faraday eq.~(\ref{Faraday}) modified in SM due to anomalies.

We generalized the approach for the magnetic field generation in a magnetar, which has been recently suggested in ref.~\cite{Dvornikov:2014uza}, where the concept of the maximum helicity density was used. In the present work we considered an arbitrary initial magnetic helicity density in eq.~(\ref{initialhelicity}), which is parametrized by the parameter $0\leq q\leq 1$. Here we also chose the continuous magnetic energy spectrum in eq.~(\ref{initial_energy}) instead of the monochromatic one in ref.~\cite{Dvornikov:2014uza}: $\rho_\mathrm{B}(k,t)=\rho_\mathrm{B}(t)\delta (k - k_0)$. The growth
of the magnetic helicity density for the initially nonhelical field, $h(k,t_0)=0$ corresponding to $q=0$, is owing to the presence of the term $\sim \rho_\mathrm{B}(k,t)$, which is nonzero at $t = t_0$, in the first line in eq.~(\ref{general}).  Although at early times nonhelical ($q=0$) and maximum helical ($q=1$) magnetic helicities are well distinguishable, the remarkable issue here is the tendency of the zero initial magnetic helicity ($q=0$) to grow up to the maximal one ($q=1$) at large time scales, compare figures~\ref{1a} and~\ref{1c} with figures~\ref{1b} and~\ref{1d}. Of course, this helicity enhancement provides the similar dependence on the parameter $q$ for the magnetic field itself in figure~\ref{fig:B}.

As shown in section~\ref{t0}, the cooling of a NS core due to the neutrino emission via the slow (modified) Urca processes leads to the increase of the electric conductivity. Such a cooling displaces with time the evolution curves for both $h(t)$ and $B(t)$ keeping the general effect of their growth at large times $\sim (10^3 - 10^5)\thinspace\text{yr}$.

Note that our approach that is based on the consideration of the binary combinations, such as $\rho_\mathrm{B}\sim B^2$ and $h\sim AB$, gives no information on the structure and the orientation of the magnetic field. We deal here with random magnetic fields having Kolmogorov's spectrum of the magnetic energy density. Such magnetic fields are characterized by the changing amplitude $B$ and the varying large spatial scale $\Lambda_\mathrm{B}=k^{-1}$ in the continuous spectrum. Decreasing the minimal scale $\Lambda_\mathrm{B}^{(\mathrm{min})}=k_\mathrm{max}^{-1}$ down to the inhomogeneity size of velocities $\lambda_v$ we should also consider the Navier-Stokes equation for the random fluid velocity ${\bf v}$ as well as distinguish a large scale magnetic field ${\bf B}$ and a small scale fluctuation ${\bf b}$. Then the interplay of the kinetic helicity $\alpha_\mathrm{kin}\sim \langle {\bf v}\cdot(\nabla\times {\bf v})\rangle$, where the mean value is taken at large scales\footnote{In the dynamo term $\nabla\times \langle {\bf v}\times {\bf b}\rangle$, the mean vector $\langle {\bf v}\times {\bf b}\rangle\approx \alpha_\mathrm{kin}{\bf B}$ defines the pseudoscalar $\alpha_\mathrm{kin}\sim \langle {\bf v}\cdot(\nabla\times {\bf v})\rangle$.}, and the anomalous magnetic helicity parameter $\alpha=\Pi/\sigma_\mathrm{cond}$ entering the Faraday eq.~(\ref{Faraday}), both governing the evolution of large scale magnetic fields, would be important. In the present approach we do not take into account these problems. It should be also noted that in the case of the solid-state rotation with the spatially homogeneous angular velocity $\Omega(t)$ we assume here, the azimuthal rotation velocity of NS  as a whole, ${\bf V}_0=\Omega (t) r\sin\theta \mathbf{e}_{\Phi}$, $0\leq r\leq R_\mathrm{NS}$, does not contribute to the dynamo term $\nabla\times({\bf V}\times {\bf B})$, where ${\bf V}={\bf V}_0 + {\bf v}$ is  the total fluid velocity and ${\bf v}$ is the small
scale (random) velocity vanishing at large scales. As a result, the $\alpha\Omega$ dynamo is not essential for us, while $\alpha^2$-dynamo is active, see, e.g., refs.~\cite{Dvornikov:2013bca,ZelRuzSok90}.

To resume we have further developed a novel mechanism, initially proposed in ref.~\cite{Dvornikov:2014uza}, for the generation of the strongest magnetic fields in magnetars taking into account: (i)~the neutrino cooling of the NS core, (ii)~continuous magnetic helicity density and magnetic energy density spectra, and assuming (iii)~an arbitrary initial magnetic helicity, including the case of the initially non-helical magnetic field. We have found that the CS anomaly caused by the parity violation in the SM interaction of electrons with nucleons drives the growth of a seed magnetic field by about five orders of magnitude and its helicity by about ten orders of magnitude independently of an initial magnetic helicity density $h(t_0)$ that could be even zero, $h(t_0)=0$.  Such a sharp magnetic helicity growth, followed by the strong enhancement of the magnetic field, is the main result of the present work.

\acknowledgments

We are thankful L.B.~Leinson and D.D.~Sokoloff for useful discussions and P.M.~Woods for communications. M.D. is grateful to FAPESP (Brazil) for the Grant No.~2011/50309-2, to the Competitiveness Improvement Program at the Tomsk State University and
to RFBR (research project No.~15-02-00293) for partial support.

\appendix

\section{Averaged electron current induced by chiral effects\label{CURRDER}}

In this appendix we obtain the solution of the Dirac equation for a massless electron interacting with nucleons under the influence of the magnetic field. Then we derive the induced electric current along the magnetic field direction.

The Dirac equation for massless electrons reads
\begin{equation}\label{eq:Direqpsie}
  \left[
    \gamma^{\mu}
    \left(
      \mathrm{i}\partial_{\mu}+eA_{\mu}
    \right) -
    \gamma^{0}
    \left(
      V_{\mathrm{L}}P_{\mathrm{L}}+V_{\mathrm{R}}P_{\mathrm{R}}
    \right)
  \right]
  \psi_{e}=0,
\end{equation}
where $\gamma^{\mu}=\left(\gamma^{0},\bm{\gamma}\right)$ are the
Dirac matrices, $A^{\mu}=\left(0,0,Bx,0\right)$ is the vector potential
for the magnetic field directed along the $z$-axis, $P_{\mathrm{L,R}}=(1\mp\gamma^{5})/2$
are the chiral projection operators, $\gamma^{5}=\mathrm{i}\gamma^{0}\gamma^{1}\gamma^{2}\gamma^{3}$, $e>0$ is the absolute value of the electron charge, and $V_{\mathrm{L,R}}$ are the effective potentials for the interaction of left and right electrons with nucleons. The explicit form of $V_{\mathrm{L,R}}$ can be found in ref.~\cite{Dvornikov:2014uza}. The upper sign both in $\mp $ and $\pm$, throughout this appendix, corresponds to the left particles.

Let us decompose $\psi_{e}$ in the chiral projections as $\psi_{e}=\psi_{\mathrm{L}}+\psi_{\mathrm{R}}$,
where $\psi_{\mathrm{L,R}}=P_{\mathrm{L,R}}\psi_{e}$.
Using the Dirac matrices in the standard representation~\cite{ItzZub80},
\begin{equation}
  \gamma^{0} =
  \left(
    \begin{array}{cc}
      1 & 0 \\
      0 & -1
    \end{array}
  \right),
  \quad
  \bm{\gamma} =
  \left(
    \begin{array}{cc}
      0 & \bm{\sigma} \\
      -\bm{\sigma} & 0
    \end{array}
  \right),
  \quad
  \gamma^{5} =
  \left(
    \begin{array}{cc}
      0 & 1 \\
      1 & 0
    \end{array}
  \right),
\end{equation}
where $\bm{\sigma}$ are the Pauli matrices, it is convenient to represent $\psi_{\mathrm{L,R}}^{\mathrm{T}} = \left( \varphi_{\mathrm{L,R}},\mp\varphi_{\mathrm{L,R}} \right)$.
Using eq.~(\ref{eq:Direqpsie}) and separating the variables $\varphi_{\mathrm{L,R}} = e^{-\mathrm{i}E_{\mathrm{L,R}}t + \mathrm{i}p_{y}y+\mathrm{i}p_{z}z}\varphi_{\mathrm{L,R}}(x)$,
we get the following equation for $\varphi_{\mathrm{L,R}}(x)$:
\begin{equation}\label{eq:phiL}
  \left[
    P_{0}\pm
    \left(
      \bm{\sigma}\mathbf{P}
    \right)
  \right]
  \varphi_{\mathrm{L,R}} =
  \left(
    \begin{array}{cc}
      P_{0} \pm p_{z} & \mp\mathrm{i}\sqrt{eB}
      \left[
        \partial_{\eta}+\eta
      \right]
      \\
      \mp\mathrm{i}\sqrt{eB}
      \left[
        \partial_{\eta}-\eta
      \right]
      &
      P_{0}\mp p_{z}
    \end{array}
  \right)
  \varphi_{\mathrm{L,R}}=0,
\end{equation}
where $P^{\mu} = \left( E_{\mathrm{L,R}}-V_{\mathrm{L,R}},-\mathrm{i}\partial_{x},p_{y}+eBx,p_{z} \right)$
and $\eta=\sqrt{eB}x+p_{y}/\sqrt{eB}$.

The solution of eq.~(\ref{eq:phiL}) can be found using the Hermite
function $u_{\mathrm{n}}(\eta)  = \left(eB/\pi\right)^{1/4} \\ \times \exp(-\eta^{2}/2)H_{\mathrm{n}}(\eta)/\sqrt{2^{\mathrm{n}}\mathrm{n}!}$,
where $H_{\mathrm{n}}(\eta)$ is the Hermite polynomial, as
\begin{align}\label{eq:phiLun}
  \varphi_{\mathrm{L,R}}(x)= & \frac{1}{4\pi\sqrt{P_{0}}}
  \left(
    \begin{array}{c}
      \sqrt{P_{0}\mp p_{z}}u_{\mathrm{n}-1}
      \\
      \mp\mathrm{i}\sqrt{P_{0} \pm p_{z}}u_{\mathrm{n}}
    \end{array}
  \right),
\end{align}
for $\mathrm{n}=1,2,\dotsc$. The normalization coefficient in eq.~(\ref{eq:phiLun})
corresponds to the following normalization of the four component wave
function $\psi_{\mathrm{L,R}}^{\mathrm{T}} = \left( \varphi_{\mathrm{L,R}},\mp \varphi_{\mathrm{L,R}} \right)$:
\begin{equation}
  \int
  \left(
    \psi_{\mathrm{L,R}}
  \right)_{\mathrm{n}p_{y}p_{z}}^{\dagger}
  \left(
    \psi_{\mathrm{L,R}}
  \right)_{\mathrm{n}'p'_{y}p'_{z}}
  \mathrm{d}^{3}x =
  \delta_{\mathrm{n}\mathrm{n}'} \delta
  \left(
    p_{y}-p'_{y}
  \right)
  \delta
  \left(
    p_{z}-p'_{z}
  \right).
\end{equation}
The energy levels can be found from the expression,
\begin{equation}\label{eq:EL}
  P_{0}^{2} =
  \left(
    E_{\mathrm{L,R}}-V_{\mathrm{L,R}}
  \right)^{2} =
  p_{z}^{2}+2eB\mathrm{n}.
\end{equation}
To obtain the solution in eq.~(\ref{eq:phiLun}) we use the following
properties of the Hermite functions: $\left[ \partial_{\eta}+\eta \right] u_{\mathrm{n}} = \sqrt{2\mathrm{n}}u_{\mathrm{n}-1}$
and $\left[ \partial_{\eta}-\eta \right] u_{\mathrm{n}-1} = -\sqrt{2\mathrm{n}}u_{\mathrm{n}}$.

If $\mathrm{n}=0$, the solution of eq.~(\ref{eq:phiL}) has the form,
\begin{equation}\label{eq:phiLu0}
  \varphi_{\mathrm{L,R}}(x) = \frac{1}{2\pi\sqrt{2}}
  \left(
    \begin{array}{c}
      0 \\
      u_{0}
    \end{array}
  \right).
\end{equation}
Substituting eq.~(\ref{eq:phiLu0}) to eq.~(\ref{eq:phiL}) and
using eq.~(\ref{eq:EL}), we get that $p_{z}>0$ for left particles and $p_{z}<0$ for the right ones. It should be noted
that, at $\mathrm{n}>0$, $-\infty<p_{z}<+\infty$.

Finally, we obtain the four component wave function in the form,
\begin{align}\label{eq:psiL}
  \psi_{\mathrm{L,R}}^{(\mathrm{n}>0)}(x)= &
  \frac{1}{4\pi\sqrt{E_{\mathrm{L,R}}-V_{\mathrm{L,R}}}}
  \left(
    \begin{array}{c}
    \sqrt{E_{\mathrm{L,R}}-V_{\mathrm{L,R}}\mp p_{z}}u_{\mathrm{n}-1}
    \\
    \mp\mathrm{i}\sqrt{E_{\mathrm{L,R}}-V_{\mathrm{L,R}}\pm p_{z}}u_{\mathrm{n}}
    \\
    \mp\sqrt{E_{\mathrm{L,R}}-V_{\mathrm{L,R}}\mp p_{z}}u_{\mathrm{n}-1}
    \\
    \mathrm{i}\sqrt{E_{\mathrm{L,R}}-V_{\mathrm{L,R}}\pm p_{z}}u_{\mathrm{n}}
  \end{array}
  \right),
  \notag
  \\
  \psi_{\mathrm{L,R}}^{(\mathrm{n}=0)}(x) = & \frac{1}{2\pi\sqrt{2}}
  \left(
    \begin{array}{c}
      0
      \\
      u_{0}
      \\
      0
      \\
      \mp u_{0}
    \end{array}
  \right).
\end{align}

Using eq.~(\ref{eq:psiL}), we can calculate the averaged electric current
along the magnetic field as
\begin{equation}\label{eq:Jzelposgen}
  J_{z}^{\mathrm{L,R}}= e
  \sum_{\mathrm{n}=0}^{\infty}
  \int_{-\infty}^{+\infty}\mathrm{d}p_{y}
  \int\mathrm{d}p_{z}
  \left[
    \bar{\psi}_{\mathrm{L,R}}^{\bar{e}} \gamma^{3} 
    \psi_{\mathrm{L,R}}^{\bar{e}}
    f_{\bar{e}}(E_{\mathrm{L,R}}^{\bar{e}}) -
    \bar{\psi}_{\mathrm{L,R}}^{e} \gamma^{3} \psi_{\mathrm{L,R}}^{e}
    f_e(E_{\mathrm{L,R}}^{e})
  \right],
\end{equation}
where $f_{e,\bar{e}}(E)=\left[\exp(\beta ( E \mp \mu_\mathrm{L,R})+1\right]^{-1}$ is the Fermi-Dirac
distribution with upper (lower) sign ahead chemical potentials for electrons (positrons), $E_{\mathrm{L,R}}^{e,\bar{e}} = \sqrt{p_{z}^{2}+2eB\mathrm{n}} \pm V_{\mathrm{L,R}}$ are the energy levels for electrons (upper sign) and positrons (lower sign),
$\beta=1/T$ is the reciprocal temperature, and $\mu_{\mathrm{L,R}}$
is the chemical potential. The positron wave functions $\psi_{\mathrm{L,R}}^{\bar{e}}$ in eq.~\eqref{eq:Jzelposgen} can be obtained from the electron ones $\psi_{\mathrm{L,R}}^{e}$, given in eq.~\eqref{eq:psiL}, by the charge conjugation~\cite{Vilenkin}.

At $\mathrm{n}>0$ we get for electrons
\begin{align}
  \bar{\psi}_{\mathrm{L,R}}\gamma^{3}\psi_{\mathrm{L,R}} = &
  \mp\frac{1}{8\pi^{2}(E_{\mathrm{L,R}}-V_{\mathrm{L,R}})}
  \notag
  \\
  & \times
  \left[
    \left(
      E_{\mathrm{L,R}}-V_{\mathrm{L,R}}\mp p_{z}
    \right)
    u_{\mathrm{n}-1}^{2} -
    \left(
      E_{\mathrm{L,R}}-V_{\mathrm{L,R}}\pm p_{z}
    \right)u_{\mathrm{n}}^{2}
  \right],
\end{align}
and
\begin{equation}\label{intpyn}
  \int_{-\infty}^{+\infty}\mathrm{d}p_{y}
  \bar{\psi}_{\mathrm{L,R}}\gamma^{3}\psi_{\mathrm{L,R}} =
  \frac{eB}{4\pi^{2}}\frac{p_{z}}{E_{\mathrm{L,R}}-V_{\mathrm{L,R}}}.
\end{equation}
Integrating eq.~\eqref{intpyn} over $p_{z}$, one obtains that Landau levels with
$\mathrm{n}>0$ do not contribute to the electric current.

For the lowest Landau level with $\mathrm{n}=0$ we have
\begin{equation}
  \bar{\psi}_{\mathrm{L,R}}\gamma^{3}\psi_{\mathrm{L,R}} =
  \pm\frac{u_{\mathrm{0}}^{2}}{4\pi^{2}},
  \quad
  \int_{-\infty}^{+\infty}\mathrm{d}p_{y}
  \bar{\psi}_{\mathrm{L,R}}\gamma^{3}\psi_{\mathrm{L,R}} =
  \pm \frac{eB}{4\pi^{2}}.
\end{equation}
Finally one obtains for left electrons
\begin{equation}
  J_{z}^{\mathrm{L}} =
  -\frac{e^{2}B}{4\pi^{2}}\int_{0}^{+\infty}\mathrm{d}p_{z}f_e(p_{z}+V_{\mathrm{L}}),
\end{equation}
and for right electrons
\begin{equation}
  J_{z}^{\mathrm{R}} =
  +\frac{e^{2}B}{4\pi^{2}}\int_{-\infty}^{0}\mathrm{d}p_{z}f_e(-p_{z}+V_{\mathrm{R}}),
\end{equation}
and the analogous expression for positrons.

Adding the contribution of positrons with the properly changed signs,
we obtain that the total current $\mathbf{J}=\mathbf{J}_{\mathrm{L}}+\mathbf{J}_{\mathrm{R}}$ induced by chiral effects reads
\begin{align}\label{eq:j5gen}
  J_z = &
  \frac{e^{2} B}{4\pi^{2}}
  \bigg\{
    \int_{-\infty}^{0}\mathrm{d}p_z
    \left[
      f_e
      \left(
        -p_z+V_{\mathrm{R}}
      \right) -
      f_{\bar{e}}
      \left(
        -p_z-V_{\mathrm{R}}
      \right)
    \right]
      \notag
      \\
      & -
    \int_{0}^{+\infty}\mathrm{d}p_z
    \left[
      f_e
      \left(
        p_z+V_{\mathrm{L}}
      \right) -
      f_{\bar{e}}
      \left(
        p_z-V_{\mathrm{L}}
      \right)
    \right]
  \bigg\},
\end{align}
or finally (see eq.~(\ref{current}) above):
\begin{equation}\label{final}
  {\bf J}=\frac{2\alpha_{\mathrm{em}}}{\pi}(\mu_{5}+V_{5}){\bf B},
\end{equation}
which is additive to the ohmic current ${\bf J}_\mathrm{Ohm}$ in a standard QED plasma and where $\alpha_{\mathrm{em}}=e^{2}/4\pi$ is the fine structure constant, $\mu_{5}=(\mu_{\mathrm{R}}-\mu_{\mathrm{L}})/2$, and $V_{5}=(V_{\mathrm{L}}-V_{\mathrm{R}})/2$.
It is worth to mention that eq.~(\ref{final}) is valid for any electron temperature. The first term ($\sim \mu_5$) in  eq.~(\ref{final}) determines CME exploited, e.g., in QCD plasma~\cite{Kharzeev}, while the second term ($\sim V_5$) is given by weak $eN$ interactions in SM and has the polarization origin (compare in ref.~\cite{Semikoz:2011tm}).

\end{document}